\documentclass[useAMS,fleqn,usenatbib]{mnras}

\usepackage{newtxtext,newtxmath}
\usepackage{epsf,rotating,url}

\usepackage[T1]{fontenc}

\DeclareRobustCommand{\VAN}[3]{#2}
\let\VANthebibliography\thebibliography
\def\thebibliography{\DeclareRobustCommand{\VAN}[3]{##3}\VANthebibliography}

\usepackage{graphicx}	
\usepackage{amsmath}	

\newcommand{\epslp}{{\epsilon_{\rm LP}}}
\newcommand{\angstrom}{\mbox{\normalfont\AA}}

\title[The Accretion Disc in 3C\,273]{Accretion Disc Reverberation Mapping of the Quasar 3C\,273}

\author[J. P. Thorne et al.]{James P. Thorne$^{1,2}$\thanks{James Thorne passed away on 2024 February 8th. This work is based largely on his results obtained as part of his MScR thesis. His passion for research will always be remembered.}, Hermine Landt$^{1}$\thanks{E-mail: hermine.landt@durham.ac.uk}, Jiamu Huang$^3$, Juan V. Hern\'andez Santisteban$^4$, Keith Horne$^4$, 
\newauthor
Edward M. Cackett$^5$, Hartmut Winkler$^6$, David Sanmartim$^{7,8}$
\\
$^1$Centre for Extragalactic Astronomy, Department of Physics, University of Durham, South Road, Durham DH1 3LE, UK \\
$^2$Department of Physics, Imperial College London, Prince Consort Road, London SW7 2BW, UK \\
$^3$Department of Physics, University of California, Santa Barbara, CA 93106, USA \\
$^4$SUPA Physics and Astronomy, University of St. Andrews, Fife, KY16 9SS, UK \\
$^4$Wayne State University, Department of Physics \& Astronomy, 666 W Hancock St, Detroit, MI 48201, USA \\
$^6$Department of Physics, University of Johannesburg, P.O. Box 524, 2006 Auckland Park, Johannesburg, South Africa \\
$^7$Carnegie Observatories, Las Campanas Observatory, Casilla 601, La Serena, Chile \\
$^8$Rubin Observatory Project Office, 950 N. Cherry Ave., Tucson, AZ 85719, USA \\
}

\date{Accepted XXX. Received YYY; in original form ZZZ}

\pubyear{2025}

\begin{document}

\def\la{\mathrel{\hbox{\rlap{\hbox{\lower4pt\hbox{$\sim$}}}\hbox{$<$}}}}
\def\ga{\mathrel{\hbox{\rlap{\hbox{\lower4pt\hbox{$\sim$}}}\hbox{$>$}}}}

\font\sevenrm=cmr7
\def\OIII{[O~{\sevenrm III}]}
\def\FeII{Fe~{\sevenrm II}}
\def\FeIIf{[Fe~{\sevenrm II}]}
\def\SIII{[S~{\sevenrm III}]}
\def\HeI{He~{\sevenrm I}}
\def\HeII{He~{\sevenrm II}}
\def\NeV{[Ne~{\sevenrm V}]}
\def\OIV{[O~{\sevenrm IV}]}

\def\iraf{{\sevenrm IRAF}}
\def\mpfit{{\sevenrm MPFIT}}
\def\galfit{{\sevenrm GALFIT}}
\def\prepspec{{\sevenrm PrepSpec}}
\def\mapspec{{\sevenrm mapspec}}
\def\cream{{\sevenrm CREAM}}
\def\javelin{{\sevenrm JAVELIN}}
\def\cloudy{{\sevenrm CLOUDY}}
\def\astroimagej{{\sevenrm AstroImageJ}}
\def\banzai{{\sevenrm BANZAI}}
\def\orac{{\sevenrm ORAC}}
\def\demc{{\sevenrm DEMC}}

\label{firstpage}
\pagerange{\pageref{firstpage}--\pageref{lastpage}}
\maketitle

\begin{abstract}
We present accretion disc size measurements for the well-known quasar 3C\,273 using reverberation mapping (RM) performed on high-cadence light-curves in seven optical filters collected with the Las Cumbres Observatory (LCO). Lag estimates obtained using \texttt{Javelin} and \texttt{PyROA} are consistent with each other and yield accretion disc sizes a factor of $\sim 2-7$ larger than `thin disc' theoretical expectations. This makes 3C\,273 one of a growing number of active galactic nuclei (AGN) to display the so-called `accretion disc size' problem usually observed in low-luminosity AGN. Power-law fits of the form $\tau \propto \lambda^{\beta}$ to the lag spectrum, and $\nu\,f_\nu\propto \nu^\beta$ to the spectral energy distribution (SED) of the variations, both give results consistent with the `thin disc' theoretical expectation of $\beta = 4/3$. The Starkey et al. `flat disc with a steep rim' model can fit both the lag estimates and the SED variations. Extrapolating the observed optical lags to putative dust-forming regions of the disc gives $r\sim 100$ -- 200\,light-days. These radii are consistent with the size of the broad line region (BLR) as determined by near-infrared interferometric studies as well as with the best-fit location of the outer edge for the `flat disc with a steep rim' model. Therefore, the accretion disc in 3C\,273 might be sufficiently extended to be dusty, allowing the BLR to emerge from it in a dusty outflow. A flux variation gradient analysis and the structure function of our LCO light-curves confirm that the optical variability in 3C\,273 is dominated by the accretion disc rather than its radio jet. 
\end{abstract}




\section{Introduction}\label{intro}

Active galactic nuclei (AGN) are compact yet highly luminous phenomena, generating optical luminosities similar to or greater than the total starlight luminosity of their host galaxies.
Quasars are the brightest subset of AGN, typically outshining the host galaxy starlight by factors of a hundred or more. The standard paradigm of quasars consists of a layered structure comprised of a central supermassive black hole (SMBH) surrounded by a luminous accretion disc followed by a broad emission line region (BLR) with a molecular dusty torus forming the outer boundary \citep[see review by][]{2015ARA&A..53..365N}. The accretion disc is of particular importance since it is expected to regulate black hole growth and provide the primary source of radiation for ionising the surrounding gas and heating the circumnuclear dust.

The main mechanism thought to generate luminosity in the accretion disc is viscous heating through mass accretion onto the SMBH \citep{Zeldovich1964}. 
The total bolometric luminosity in the disc 
is then
\begin{eqnarray}
L_{\rm bol} = \eta\, \dot{M}\, c^{2},
\label{eq:bolo}
\end{eqnarray}

\noindent
where $\dot{M}$ is the mass accretion rate, $c$ is the speed of light, and $\eta$ is the efficiency of energy conversion (typically assumed as $\eta \sim 0.1$). 
The standard method used to model accretion disc emission follows the approach of \citet{Shakura1973} by prescribing a geometrically thin, optically thick disc. In this regime the disc annuli can be described as separate blackbodies with emission given by Planck's distribution and controlled by a temperature gradient across the disc of the form
\begin{eqnarray}
    T^{4}(r) = 
    \frac{3\,G\,M\,\dot{M}}{8\,\pi\, \sigma\, r^{3}} 
    \left(1 - \sqrt{\frac{r_{\rm isco}}{r}} \right) \ ,
\label{eq:temp-radius}
\end{eqnarray}
\noindent
where $M$ is the black hole mass, $\sigma$ is the Stefan-Boltzmann constant and $r_{\rm isco}$ is the radius of innermost stable circular orbit around the SMBH. The factor $3\,(1 - \sqrt{r_{\rm isco}/r})$ accounts for radial re-distribution of the gravitational energy released in the disc due to the work done by viscous torques, assuming zero torque at the inner edge of the disc at $r_{\rm isco}$. This approaches 3 at $r \gg r_{\rm isco}$. 
Combining Eq.\,(\ref{eq:temp-radius}) with Wien's displacement law ($T \propto \lambda_{\rm max}^{-1}$) gives an expectation for the peak wavelength emitted by each disc annulus to scale with radius as $r \propto \lambda^{\beta}$. 
Integrating the blackbody emission of the annuli over the entire disc gives a similar power-law expectation for the flux, $\nu f_{\nu}\propto \nu^{\beta}$. 
In both cases, $T\propto r^{3/4}$ corresponds to $\beta = 4/3$ according to the \citet{Shakura1973} model.

It is not currently possible to spatially resolve the inner regions of most AGN owing to the small angular sizes involved. 
Reverberation Mapping (RM) is an indirect method of probing the structure of AGN accretion flows, using light travel time delays to measure distances between the compact irradiating source near the SMBH and the surrounding structures, including the BLR and torus as reprocessed accretion disc flux \citep{Cackett_2021}. 
Determining the response time (or lag) $\tau$ between the observed light-curves from the BLR and torus with respect to the accretion disc continuum allows estimates on the distances, $r$, between these structures to be made. This approach assumes that light-travel time is the most significant timescale, meaning that the lag can be described as $\tau(r) = r/c$. Within the accretion disc, continuum photometric light-curves are observed to lag with respect to light-curves from shorter wavelength bands \citep[e.g.][]{2007MNRAS.380..669C,Edelson_2019,Jha_2022}. RM between different continuum light-curves can be used to estimate distances within the accretion disc between the different annuli emitting predominantly at these wavelengths \citep{Cackett_2021}. The power laws introduced above lead to the expectation that lags within the accretion disc will scale with wavelength according to $\tau \propto \lambda^{\beta}$ where $\beta = 4/3$. 

The standard explanation for the observed variability in accretion disc light-curves is through irradiation of the disc surface by a hypothesised `lamp-post' structure, modelled as a compact emitter of highly variable X-rays situated above and below the centre of the disc \citep{Collin_2003}. 
In this picture, the rapidly variable lamp-post flux is reprocessed by each of the blackbody annuli in the disc and re-emitted at longer wavelengths. The temporal structure of the variability is thus preserved, but with a delay and a degree of blurring due to the light travel times. 
The cooler outer disc emits longer-wavelength reprocessed light-curves with a greater lag than the inner disc due to the greater distance between the lamp-post and these regions. 

Recent studies have revealed problems with the traditional lamp-post description. This includes analysis of observed light-curves by \citet{Burke_2021} and \citet{2017MNRAS.470.3591G} that find characteristic AGN variability time-scales much longer than the rapid variability expected from reprocessed X-ray emission. In fact, \citet{Burke_2021} and \citet{Kelly_2009} find that the variability time-scales are in much better agreement with expected thermal time-scales. These findings may imply that flux variations in the disc arise from local random temperature perturbations rather than from reprocessed X-ray flux. In addition, accretion disc RM of NGC\,5548 light-curves \citep{2017FrASS...4...55F} carried out using the \texttt{PyceCREAM} fitting algorithm \citep{2016MNRAS.456.1960S} revealed a poor match between the derived model driving light-curve and the observed hard and soft X-ray light-curve data. This suggests X-rays may not be generating the optical light-curves emitted in this accretion disc. Also, \citet{Breedt_2009} found for Mrk~79 that optical variations on timescales of months are well reproduced by a reprocessing model, but longer term variations of years cannot be accounted for in this way. These results highlight the need to modify the standard `lamp-post' description. 

Accretion disc RM faces unique observational challenges compared to RM performed on larger length scales. In particular, AGN light-curves typically show relatively small amplitude variability on the time-scales corresponding to the shorter expected lags \citep[e.g.][]{Mudd2018}. This makes it difficult to constrain time lags unless sufficiently high cadence and high precision light curve data are available. Despite these challenges, previous accretion disc RM studies do yield some consistent findings. The correlation between increasing light-curve lags with longer wavelengths is well established, and the expected $\tau\propto\lambda^{4/3}$ power law noted above is generally consistent with the observed lags \citep[e.g.][]{2007MNRAS.380..669C,2015ApJ...806..129E}. However much still remains unknown about the accretion discs of AGN, motivating further study of these structures. 

A major puzzle of current interest, referred to by \citet{Cackett_2021}
as the `accretion disc size problem', 
concerns the inferred sizes of accretion discs being consistently larger, typically by a factor $\sim 2-3$, than expected from the standard thin-disc model \citep[e.g.][]{McHardy_2014,2015ApJ...806..129E,2016ApJ...821...56F,Jiang_2017,Mudd2018,2018ApJ...857...53C,Edelson_2019,Cackett_2020,JVHS2020}. 
Measuring disc sizes for AGN samples that span a wide range in mass and luminosity is needed to establish and test predicted scaling relations. 
The results should help us to understand whether the disc size issue represents a problem with the underlying accretion theory or with certain approximations made in measuring observed lags
and calculating the corresponding predicted lags. 
These considerations motivate our efforts to study bright nearby sources like 3C\,273 which are predicted to have large discs. 
This should also help to establish whether AGN accretion discs extend out to dust-forming regions, and thus to test dusty outflow formation models for the BLR \citep{Czerny2011} as we attempt to do in Section~\ref{frado}.      

Monitoring of 3C\,273 has been previously undertaken \citep[e.g.][]{Zhang_2019,fernandes2020,Figaredo_2020}, but to our knowledge no dedicated accretion disc RM campaign has so far been carried out. 
The close proximity and large luminosity in this quasar make it an ideal candidate for such an analysis. 
In Section~\ref{target}, we introduce and discuss the science target. 
Section~\ref{Data} gives the details of the observations. 
In Section~\ref{Results}, we outline the methods used to obtain the lag estimates and the theoretical predictions and present our results, which are discussed in the context of accretion disc models in Section~\ref{Discussion}. 
Finally, in Section~\ref{Conclusion}, we present a short summary and our conclusions. We assume throughout cosmological parameters $H_{0}=70$~km~s$^{-1}$~Mpc$^{-1}$, $\Omega_{\rm M} = 0.3$ and $\Omega_{\Lambda} = 0.7$.

\section{The science target} \label{target}

3C\,273 was the first discovered quasar, as identified by \citet{Schmidt1963}. Its redshift of $z=0.158$ and corresponding luminosity distance $D_{\rm L}=757$\,Mpc make it the brightest nearby AGN. 3C\,273 is large and luminous enough that direct imaging of the BLR and torus structures were recently carried out by the near-infrared (near-IR) interferometric instrument GRAVITY on ESO’s VLT \citep{Gravity2018}. These observations yielded a BLR radius of $\sim 150$~light-days, an extent of the hot dust in the torus of $\sim 900$~light-days, and a black hole mass estimate of 
$M_{\rm BH} = (2.6 \pm 1.1) \times 10^{8} M_{\odot}$.
For this black hole mass of $M_{\rm BH} \sim 3 \times 10^{8} M_{\odot}$, the corresponding Eddington luminosity is $L_{\rm Edd} = 3.8 \times 10^{46}$~erg~s$^{-1}$. An analysis of contemporaneous optical and near-IR spectroscopy of 3C\,273 carried out by \citet{Landt2011} gave estimates for the accretion disc bolometric luminosity of $L_{\rm bol} \sim 10^{47}$~erg~s$^{-1}$. 3C\,273 has a 1-sided radio jet inclined by $\approx5^\circ$ from our line of sight, with a bulk velocity $v/c>0.55$ and kinetic power $\sim10^{44}$\,erg\,s$^{-1}$ \citep[e.g.][]{Harwood_2022}, which makes it a member of the `blazar' AGN class. The significance of this radio jet on our analysis is discussed in Section~\ref{FVG and jet}. The length scales of the BLR and torus in 3C\,273 are also constrained by optical RM analysis carried out by \citet{Zhang_2019} and near-IR photometric RM carried out by \citet{Figaredo_2020}, respectively. The latter found a significantly smaller radius of the hot dust of $\sim 400$~light-days from a five-year long monitoring campaign.

\section{The Data}

\label{Data}

\begin{figure*}
	\includegraphics[width=0.9\linewidth]{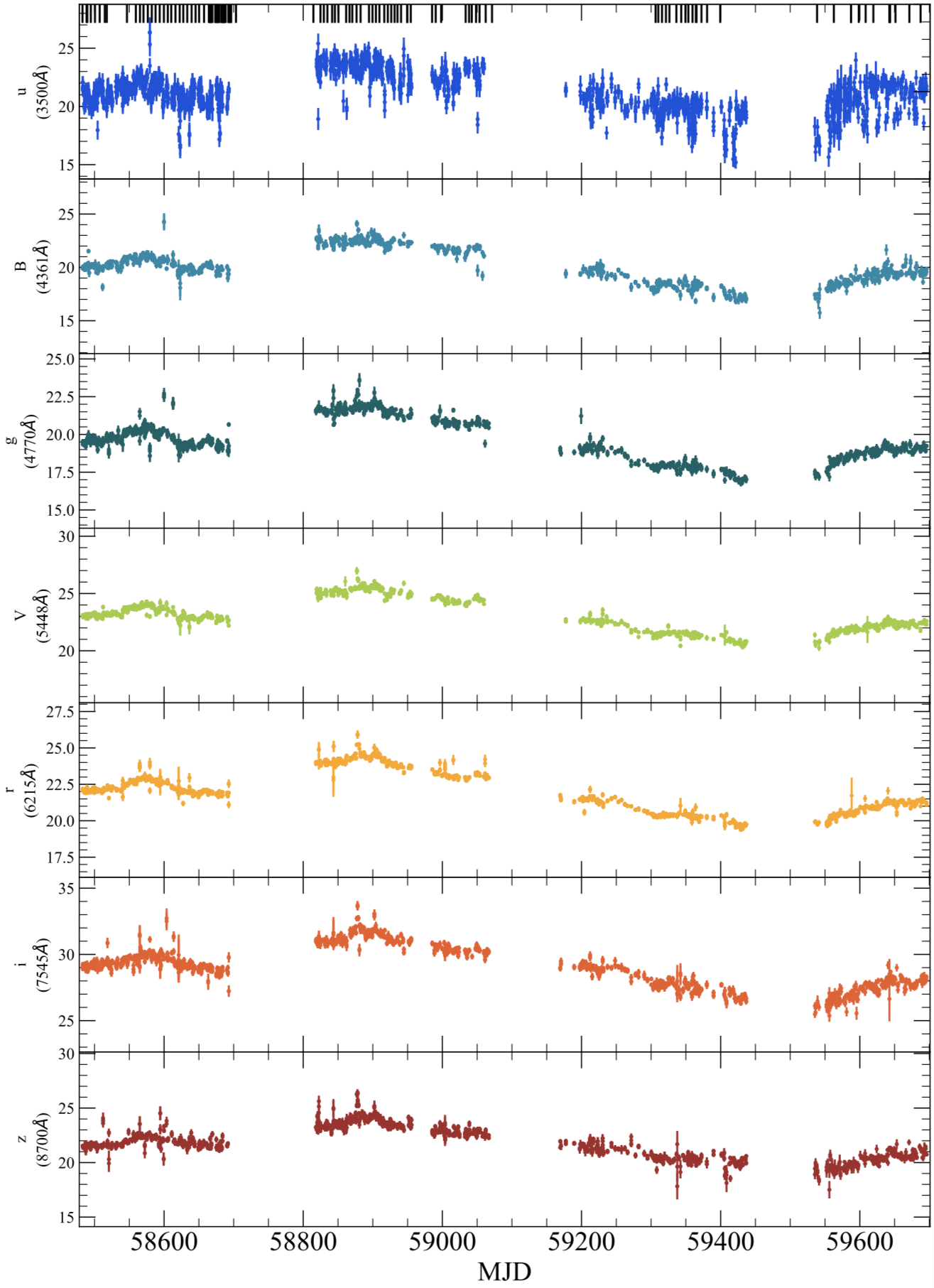}
    \caption{Multi-band LCO light-curves of 3C\,273 between 2019~Jan and 2022~Apr. Flux units are in mJy and time is represented as modified Julian date.}
    \label{fig:lc}
\end{figure*}

The data presented here comes from observations of 3C\,273 with the Las Cumbres Observatory \citep[LCO;][]{2013PASP..125.1031B} over a total period of $\sim 3.3$~years (2019~Jan -- 2022~Apr), with continuous monitoring during four observing seasons separated by gaps of $\sim 3$~months when the source is unobservable (Fig.\,\ref{fig:lc}). 
This LCO monitoring campaign was conducted in support of a long-term near-IR spectroscopic reverberation mapping campaign started in early 2019. 
LCO is a global network of robotic telescopes providing almost continuous monitoring which makes it uniquely well suited for time domain astronomy campaigns. 
Visiting 3C\,273 every 1 to 4 days, 
the Sinistro cameras on LCO's network of 1~m telescopes 
collected CCD images in each of 7 filters, the Johnson-Cousins Bessell $B$ and $V$ filters, the SDSS $u$, $g$, $r$, $i$ filters and the Pan-STARRS $z_s$ filter, hereafter referred to as \emph{uBgVriz}. 
As indicated in Table.\,\ref{tab:1}, these filters span the near-UV to the near-IR over an observer wavelength range of $\sim 3500 - 8700$~\AA. 
Exposures in each filter were taken in pairs to mitigate the impact of cosmic ray hits and to provide a check on the uncertainty estimates.
The observing period spanned 1213~days, making it much longer than the expected accretion disc time lags, which are on the order of days to weeks (see Section~\ref{theory}). 

The data were reduced and flux-calibrated as described in \citet{JVHS2020}. In short, the frames were first processed by LCO's \mbox{\banzai}~pipeline \citep{banzai} in the usual way (bias and dark subtraction, flat-fielding correction and cosmic ray rejection) and were subsequently analysed with the custom-made pipeline described in detail by \citet{JVHS2020}. After performing aperture photometry with a diameter of $7\arcsec$ and subtracting a background model, stable light-curves were produced by constructing a curve of growth using the standard stars on each individual frame and measuring the correction factors required to bring all different light-curves to a common flux level. A colour-correction and correction for atmospheric extinction were applied before the photometric calibration. Finally, an image zero-point calibration was performed at each epoch based on comparison stars in the field. 

Note in Fig.\,\ref{fig:lc} the larger errors and much poorer cross-telescope intercalibration for the $u$-band data.
Because our standard data reduction pipeline derives an absolute photometric calibration independently for each CCD image, the resulting light curve can be degraded when the field has too few suitable $u$-band calibration stars.
For this reason, we have re-extracted the $u$-band light-curve using \mbox{\astroimagej}
~\citep{Collins2017}. The LCO images are first processed by LCO's \mbox{\banzai}~pipeline \citep{banzai}. Then we use a fixed aperture of $16\arcsec$ radius to perform the differential photometry for the $u$-band data relative to 7 reference stars. The full width half maxima (FWHM) of the point-spread-functions measured from 3C\,273 and from the reference stars are $\sim 4\arcsec$ to $6\arcsec$. The differential flux is computed by dividing 3C\,273's net integrated counts by the sum of the net integrated counts of all 7 comparison stars. Then, the normalization of the lightcurve as a whole is determined from the photometric data from the comparison stars.

\begin{table}
\begin{center}
\caption{Summary of the properties of the 3C\,273 light-curve data
\label{tab:1} }
\begin{tabular}{ccccccc} 
 \hline
 Filter & $\lambda_{\rm eff}$ & FWHM & Epochs & Cadence & Broad & Line flux 
 \\
 & (\AA) & (\AA) & & (days) & Line & (per cent)
 \\
 (1) & (2) & (3) & (4) & (5) & (6) & (7)
 \\ \hline\hline
 $u$   & 3540 &  570 & 1816 & 0.480 & & 
 \\ 
 $B$   & 4361 &  890 & 1884 & 0.463 & H$\delta$ & 7 
 \\ 
 $g$   & 4770 & 1500 & 2213 & 0.401 & H$\gamma$ & 8 \\ 
 $V$   & 5448 &  840 & 1824 & 0.478 & H$\beta$, H$\gamma$ & 20 
 \\ 
 $r$   & 6215 & 1390 & 2145 & 0.414 & H$\beta$ & 13 
 \\ 
 $i$   & 7545 & 1290 & 2126 & 0.417 & H$\alpha$ & 66 
 \\ 
 $z_s$ & 8700 & 1040 & 2385 & 0.371 & & 
 \\ \hline
\end{tabular}

\parbox{85mm}{The columns are: (1) filter name; (2) filter central wavelength\footnotemark[1]; (3) filter width\footnotemark[1]; (4) total number of observations; (5) mean time between consecutive images, excluding the seasonal gaps; (6) broad emission line within the filter passband; and (7) broad line flux as a percentage of the total flux in the filter passband.} 

\end{center}
\end{table}

\begin{figure}
    \centering 
    \includegraphics[width=\columnwidth]{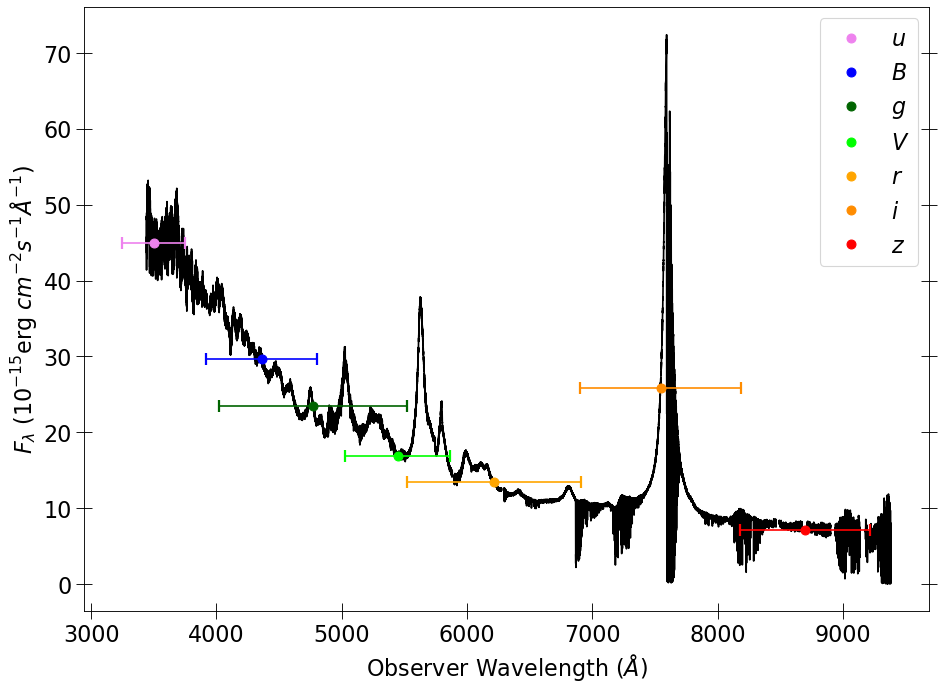}
    \caption{Contemporaneous optical spectrum of 3C\,273 obtained with the high-resolution echelle spectrograph MIKE. Overplotted are the photometric filter centres and widths used to obtain our light-curve data.
    \label{fig:mike}
    }
\end{figure}

The LCO filter passbands\footnote{\url{https://lco.global}} can include strong broad emission lines whose flux is expected to be variable, but on longer timescales than the accretion disc continuum. We have estimated the contribution from the BLR using a contemporaneous optical spectrum obtained on 2021 May 31 with the high-resolution echelle spectrograph MIKE on the 6.5~m Magellan Clay telescope, Chile. The optical spectrum obtained is shown in Fig.\,\ref{fig:mike} with the LCO filter wavelength centroids and widths included to give an indication of the parts of the spectrum sampled by our light-curves. Table~\ref{tab:1} shows the estimated contamination levels of the total flux within each passband by the broad emission lines. These were determined by fitting accretion disc power laws to the spectrum. The strong telluric absorption lines evident in Fig.\,\ref{fig:mike} were removed before determining the broad line contamination estimates. The estimates in Table~\ref{tab:1} reveal significant contamination of the $i$ band by the strong H$\alpha$ line evident in Fig.\,\ref{fig:mike}. Moderate levels of contamination occur also in the $V$ band from a combination of the H$\beta$ and H$\gamma$ lines and in the $r$ band from the strong H$\beta$ line alone. Little significant broad-line contamination occurs for the remaining photometric filters. The presence of broad emission-line flux in our light-curves will increase the differential lags because the BLR extends to larger distances from the SMBH than does the accretion disc. The presence of this biasing effect on our RM lag estimates is discussed further in Section~\ref{Discussion}.

\section{Time-series Analysis and Results} \label{Results}

Reverberation mapping (RM) techniques aim to estimate time lags between light-curves by fitting common features between them. 
A notable difficulty is the incomplete and noisy sampling of the light-curve flux data.
Time lags can be estimated with cross-correlation methods, and more detailed information can be extracted by modelling the light-curve data with more care.
Because the ionising X-ray and EUV flux variations are generally not accessible, an observed continuum light curve $f_{\rm C}(t)$ serves as a proxy.
The lagged light-curve $f_{L}(t)$ is then represented with a linearised echo model:
\begin{eqnarray}
f_{\rm L}(t) = \bar{f}_{\rm L} + 
\int \psi(\tau) \left( f_{\rm C}(t-\tau) - \bar{f}_{\rm C} \right) \, d\tau.
\label{eq:transfer-fnct}
\end{eqnarray}
\noindent
Here the reference level $\bar{f}_{\rm C}$, typically a mean or median of the continuum light curve, corresponds to a mean flux $\bar{f}_{\rm L}$ in the lagged echo light curve.
The `transfer function' $\psi (\tau)$, a distribution over time delay $\tau$, describes the marginal response of the reprocessed light at time $t$ to changes in the driver light curve at the earlier times $t-\tau$. 
With time delays arising from light travel time, the delay distribution $\psi(\tau)$ encodes information on the size and structure of the reprocessing region.

For accretion disc RM, the inclination of the disc has an important effect on the shape of the transfer function. 
According to \citet{Starkey2016}, inclination affects the shape of $\psi(\tau)$, but does not change the mean lag $\langle\tau\rangle$, for which a face-on disc may be assumed. 
In this section, we apply RM techniques to our LCO light-curve data. 
We first derive theoretical lag predictions and then measure inter-band continuum lags from the light-curves using two different RM algorithms: \texttt{Javelin} \citep{Zu2011} and \texttt{PyROA} \citep{2021MNRAS.508.5449D}. 
Finally, we use the flux variation gradient (FVG) method to derive the spectral shapes of the constant and variable components.

\subsection{Theoretical Lag Predictions} \label{theory}

Theoretical lag predictions for the LCO passbands are obtained using the \citet{Shakura1973} thin disc model. For a power-law temperature profile, the predicted lag at wavelength $\lambda$ relative to the lag at a reference wavelength $\lambda_0$, i.e. the differential lag, is:

\begin{equation}
    \Delta \tau(\lambda) = \tau_0 \left [ \left ( \frac{\lambda}{\lambda_0} \right )^\beta - 1 \right ] 
\end{equation}

\noindent
The standard disc model, with $T\propto  r^{-3/4} \propto \lambda^{-1}$, predicts $\beta=4/3$ and, following \citet{2016ApJ...821...56F, Fausnaugh2018} \citep[but see also][and references therein]{Kokubo2018, Mudd2018}, $\tau_0$ scales as:

\begin{equation}
    \left ( \frac{\tau_0}{1.0~\rm{days}} \right ) = \left ( \frac{X}{2.49} \right )^{4/3} \left ( \frac{\lambda_0}{4800~{\angstrom}} \right )^{4/3} \left ( \frac{M_{\rm BH}}{10^8~M_\odot} \right )^{2/3} \left ( \frac{\dot{m}_{\rm Edd}}{0.09} \right )^{1/3},
    \label{reflagscaling}
\end{equation}


\noindent
where $X$ is a dimensionless Wien factor, which is discussed further below, and an efficiency of 0.1 is assumed for the conversion of mass to energy during the accretion process.
For 3C\,273, we adopt the SMBH mass estimate $M_{\rm BH}=3\times10^8$\,M$_\odot$ as measured by \citet{Gravity2018}. 
We then estimate $L_{\rm acc}$ by integrating under the accretion disc continuum model obtained by fitting the continuum in the MIKE spectrum shown in Fig.\,\ref{fig:mike}. 
The resulting estimate of $L_{\rm acc}=4.5\times10^{46}$ \,erg\,s$^{-1}$ is in good agreement with the value of $L_{\rm bol}\approx10^{47}$\,erg\,s$^{-1}$ determined by \citet{Landt2011}. Our estimate implies $\dot{m}_{\rm Edd} = L_{\rm acc}/L_{\rm Edd} = 1.2$. We note that \citet{Li2022} have recently redetermined the black hole mass in 3C~273 to $M_{\rm BH}=1.15\times10^9$\,M$_\odot$. However, this increase in black hole mass leads to a higher Eddington luminosity and so to a decrease in the dimensionless mass accretion rate. Since the two changes roughly cancel out, a change in assumed black hole mass would have little impact on our result from Eq.~\ref{reflagscaling}.

In the above, the specific connection between rest-frame wavelength $\lambda$ and disc temperature $T(r)$, and thus radius $r$, is expressed by the dimensionless Wien factor:

\begin{eqnarray}
    X = \frac{h\,c}{k_{B}\,T(r)\,\lambda}
    \ ,
\label{eq:wiens}
\end{eqnarray}

\noindent
where $h$ is Planck's constant, $k_{B}$ is Boltzmann's constant.
The dimensionless Wien factor $X$  accounts for the response at wavelength $\lambda$ being distributed over a range of radii in the accretion disc. For $X=1$, the flux at wavelength $\lambda$ is from the disc annulus where the temperature is $T=h\,c/k_B\,\lambda$. A traditional value $X=4.96$ corresponds to Wien's law for the peak of the blackbody distribution \citep[see e.g.][]{1997iagn.book.....P}.
In fact,
each annulus emits a blackbody 
spectrum that contributes at all wavelengths, and thus all radii contribute to the flux at wavelength $\lambda$. In the inner disc, where higher temperatures place $\lambda$ on the Rayleigh-Jeans tail of the blackbody, the surface brightness, $B_\nu \approx 2\,k_B\,T/\lambda^2$, scales with $T\propto r^{-3/4}$. At lower temperatures, the Wien tail gives an exponential drop in surface brightness at larger radii. A flux-weighted radius, corresponding to $X=2.49$, is often adopted, following the \citet{2016ApJ...821...56F} analysis of disc lags in NGC\,5548.

\citet{Mudd2018}, in their accretion disc RM analysis of 15 `Dark Energy Survey' quasars, experimented with deriving wavelength-dependent $X$ values tailored to specific AGN.
They defined a dimensionless ratio between disc radius estimates for each wavelength (see their eqs. 5 and 7).
The estimated $X_{\rm Mudd}$ values for the LCO passbands are given in Table~\ref{table:X}. 
The slight decrease in $X$ 
with increasing wavelength can be understood as
the longer wavelength filters having 
smaller relative contributions from the inner regions of the accretion disc where the temperature decreases inward.
Overall, the estimated $X_{\rm Mudd}$ values are similar to the value of $X=2.49$ adopted by \citet{2016ApJ...821...56F}. 
Therefore, we have derived lag predictions for 3C\,273 using both $X=4.96$ and $X=2.49$, which, for a reference wavelength of $\lambda_0 = 4361/(1+z)~\angstrom$, i.e. the rest-frame $B$-band effective wavelength, results in $\tau_0 = 8.95$~days and $\tau_0 = 3.57$~days, respectively.
The two set of lag predictions (in the observed frame) are listed in the third and fourth columns of Table~\ref{table:lags}.

\begin{table}
\begin{center}
\caption{Scale factors $X$ for Wien's law, Eq.\,(\ref{eq:wiens}), derived following the approach of \citet{Mudd2018}, for each of the LCO filters
\label{table:X}}
\begin{tabular}{cccccccc} 
 \hline
 Filter & $u$ & $B$ & $g$ & $V$ & $r$ & $i$ & $z$
 \\ 
$X_{\rm Mudd}$ & 2.54 & 2.53 & 2.52 & 2.52 & 2.51 & 2.51 & 2.50
 \\ 
 \hline
\end{tabular}
\end{center}
\end{table}

\subsection{Inter-band continuum lags}

The large annual data gaps in the light-curves in Fig.\,\ref{fig:lc} (of $\sim3$\,months each) pose a problem for the light-curve fitting software as they comprise a significant portion of the total observing period. We therefore opted to remove the gaps from the analysis by analysing the four light-curve seasons separately. This follows the approach of \citet{2014ApJ...788..159K} who dealt with similarly large interruptions ($\ga 100$\,days) in their dust RM campaign. The lag estimates obtained by \texttt{Javelin} and \texttt{PyROA} for each of these observing seasons are shown in Table~\ref{table:A1}. These are differential lag estimates made with respect to the $B$-band light-curve which, as the shortest wavelength light-curve used, is assumed to be the driving light-curve by the algorithms. The $u$-band data are excluded from the lag analysis due to problematic fits obtained during preliminary investigations. This is likely a result of the excessive noise in the data-set (Fig.\,\ref{fig:lc}). We also note that, in any case, the $u$ band covers in 3C\,273 the rest-frame wavelength of the BLR diffuse continuum near the Balmer edge \citep[see, e.g., Fig.\,6 in][]{JVHS2020}, which is often invoked to explain an apparently enhanced $u$-band lag in AGN \citep{2015ApJ...806..129E}. However, we do consider the $u$-band data in our flux-flux and structure function analyses (Sections~\ref{FVG and jet} and \ref{structure}, respectively).

The 3C\,273 light-curves observed in our campaign show evidence of a long-timescale trend (Fig.\,\ref{fig:lc}). Trends in 3C\,273 optical light-curves have been detected by previous campaigns \citep[e.g.][]{Zhang_2019} and have an unknown origin. This poses potential problems for RM campaigns as the trend could introduce a new component of variability which could bias lag estimates if not corrected for \citep[][]{2010ApJ...721..715D,2014ApJ...795..149P}. To test the impact of this trend, RM analysis was carried out on the Year\,3 results with the slope both present and removed. The lag results are shown in Table~\ref{table:trend} and indicate that the trend has no significant effect on the lag estimates within $\sim 1.3\sigma$ . Nevertheless, we take the cautious approach of using the detrended Year\,3 results to perform the rest of the analysis to ensure any undetected trending effects are removed.

\begin{table*}
\begin{center}
\caption{Lags relative to the $B$ band for each photometric filter across all four observing seasons from \texttt{PyROA} and \texttt{Javelin} fits to the light-curve data shown in Fig.\,\ref{fig:lc}. 
All lags are in days in observed frame. 
}
\label{table:A1}
\renewcommand{\arraystretch}{1.5}
\begin{tabular*}{\linewidth}{@{\extracolsep{\fill}}|c|cccc|cccc|} 
 \hline 
  Filter & \multicolumn{4}{c|}{\texttt{PyROA} lag estimates} & \multicolumn{4}{c|}{\texttt{Javelin} lag estimates} 
 \\  \cline{2-9}
 & \emph{Year\,1} & \emph{Year\,2} & \emph{Year\,3} & \emph{Year\,4} & \emph{Year\,1} & \emph{Year\,2} & \emph{Year\,3} & \emph{Year\,4}
 \\ \hline
 $B$ & 0 & 0 & 0 & 0 & 0 & 0 & 0 & 0
 \\
  $g$ & $1.9^{+1.0}_{-1.0}$ & $1.3^{+1.1}_{-1.0}$ & $4.0^{+1.8}_{-2.0}$ & $52.9^{+4.1}_{-6.0}$ & $3.6^{+1.8}_{-1.6}$ & $4.2^{+1.7}_{-1.9}$ & $3.2^{+0.4}_{-0.6}$ & $3.9^{+2.0}_{-1.5}$ 
 \\
 $V$ & $3.0^{+1.1}_{-1.0}$ & $2.3^{+2.0}_{-1.6}$ & $7.2^{+1.8}_{-1.8}$ & $62.3^{+5.3}_{-13.7}$ & $10.1^{+4.4}_{-4.4}$ & $9.8^{+3.4}_{-3.4}$ & $6.9^{+1.5}_{-3.3}$ & $10.3^{+4.3}_{-5.3}$
 \\
 $r$ & $5.0^{+2.9}_{-0.9}$ & $2.3^{+6.2}_{-1.4}$ & $15.0^{+1.7}_{-2.1}$ & $92.2^{+3.7}_{-6.1}$ & $20.4^{+7.9}_{-9.0}$ & $17.8^{+7.9}_{-8.3}$ & $16.0^{+3.3}_{-9.5}$ & $15.3^{+6.8}_{-6.1}$ 
 \\
 
 $i$ & $8.1^{+5.4}_{-1.1}$ & $6.3^{+11.3}_{-1.6}$ & $17.9^{+2.9}_{-2.4}$ & $94.6^{+3.5}_{-7.3}$ & $32^{+17}_{-16}$ & $36^{+10}_{-12}$ & $15^{+15}_{-8}$ & $30^{+21}_{-12}$ 
 \\
 $z_s$ & $9.7^{+4.8}_{-1.5}$ & $6.2^{+12.8}_{-1.5}$ & $33.2^{+2.1}_{-2.4}$ & $99.0^{+0.8}_{-3.0}$ & $47^{+21}_{-20}$ & $39^{+17}_{-16}$ & $39.7^{+3.1}_{-6.4}$ & $51^{+20}_{-21}$ 
 \\ \hline
\end{tabular*}
\end{center}
\end{table*}

\subsubsection{ \texttt{Javelin} modelling}

\texttt{Javelin} \citep{Zu2011} models the (mean-subtracted) driver light-curve $X(t)$ as a damped random walk (DRW), which has an exponential autocovariance function between times $t_{i}$ and $t_{j}$ of the form 
\begin{eqnarray}
\langle X(t_{i})\,X(t_{j})\rangle = 
\sigma^2\, \exp{\left(-|t_{i}-t_{j}|/\tau_{d}\right)}
\ ,
\end{eqnarray}
\noindent
with parameters $\tau_{d}$ the decorrelation timescale and $\sigma$ the rms amplitude of long-timescale variations. 
\texttt{Javelin} estimates lags by representing lagged light-curves using Eq.\,(\ref{eq:transfer-fnct}) with a top-hat transfer function,
\begin{eqnarray}
\Psi(\tau) = \left\{
\begin{array}{rl}
A/(\tau_{2}-\tau_{1}) 
\ , &  
\tau_{1}\leq \tau \leq \tau_{2} 
\\  0  & {\rm otherwise}
\end{array}
\ . \right.
\end{eqnarray}
\noindent
Here $A$ is the amplitude of the top-hat response, the mean time lag is $\bar{\tau}\equiv(\tau_2+\tau_1)/2$ and the width $\Delta\,\tau \equiv \tau_{2}-\tau_{1}$ controls the degree of blurring in the echo light-curve. \texttt{Javelin} uses a Gaussian Process \citep[see e.g.][]{Wang2020} to fit the driving light-curve data. An initial fit to the designated driving light-curve alone establishes log-gaussian priors for $\tau_{d}$ and $\hat{\sigma}\equiv \sigma^2 \tau_d/2$. A joint analysis of $N$ echo light-curves, after performing linear detrending on the data, then uses the MCMC sampler \texttt{emcee} \citep{2013PASP..125..306F} to explore the $2+3\,N$-dimensional parameter space spanned by $\tau_{d}$ and $\hat{\sigma}^2$ for the driving light curve, and $\bar{\tau}_{i}$, $\Delta\tau_{i}$ and $A_{i}$ for the $N$ echo light-curves.

We ran the \emph{`Rmap'} mode of \texttt{Javelin} on the LCO light-curves, confining lags to a range between $0$ and $4$ times the predicted theoretical lags (see Section~\ref{theory}). This approach accommodates the potential emergence of the `accretion disc size problem', which can increase lags by a factor $\sim 2-3$ above their expected values. Following the approach of \citet{Mudd2018}, we also restricted the decorrelation timescale to $50<\tau_d<300$\,days, making it consistent with the typical decorrelation timescales found in AGN by \citet{Macleod_2010} using the Sloan Digital Sky Survey. These parameter restrictions  were necessary to help constrain the fits due to the large number of parameters introduced by the six LCO light-curves. 
The final analysis employed 100 MCMC walkers each taking 
1000 steps to provide 
$10^5$ MCMC samples. 
The resulting posterior parameter distributions are shown in the Appendix Fig.\,\ref{fig:jav_par}. The optimised parameter values are taken to be the median of the MCMC parameter samples, with upper and lower uncertainties at the $84^{\mbox{th}}$ and $16^{\mbox{th}}$ percentiles, respectively. The lag results for each season are shown in Table~\ref{table:A1} and in Fig.\,\ref{fig:tau-lam}. The \texttt{Javelin} results show good consistency across all seasons within the $1\sigma$ uncertainties. 

\texttt{Javelin} does not take into account that the errors on the data might be underestimated. However, as \citet{Yu2020} discussed in detail, whereas an increase of the error bars leads to an increase in the lag uncertainty, the lag values are largely unaffected since the scaling of the errors does not change the `shape' of the light-curve.

\subsubsection{\texttt{PyROA} modelling}

\texttt{PyROA} \citep{2021MNRAS.508.5449D} models the light-curves using a running optimal
average (ROA) for the light curve shape. 
The ROA at time $t$ is defined as an optimal (inverse-variance weighted) average over the flux data,
\begin{eqnarray}
X(t) = \frac{\sum_{i=1}^{N}D_{i}\, W_i(t)
}{\sum_{i=1}^{N} W_i(t)
}
\ , \quad W_i(t)\equiv \frac{1}{\sigma_i^2}\,\exp{\left(-\frac{(t-t_i)^2}{2\,\Delta^2}\right)}
\ ,
\label{eq:roa}
\end{eqnarray}

\noindent
where $D_{i}$ are the $N$ flux data (with error-bars $\sigma_{i}$) measured at times $t_{i}$. 
This approach follows a traditional inverse variance weighted optimal average, but with a Gaussian window that lessens the influence of data points according to how far they are in time. The ROA parameter $\Delta$ parameter controls its flexibility. 
If $\Delta$ is small, $X(t)$ can follow rapid variations in the data, leading to a close fit. If $\Delta$ is large, $X(t)$ can only follow slower variations in the data, leading to a smoother fit. 

Our \texttt{PyROA} fits to $N$ light-curves use Eq.\,(\ref{eq:transfer-fnct}) 
with a sharp delay distribution, $\psi(\tau)\propto\delta(\tau-\tau(\lambda))$.
With the ROA light-curve shape $X(t)$ normalised to a mean of 0 and rms of 1, the echo light-curve model at wavelength $\lambda$ is
\begin{equation} \label{eq:fbax}
F(t,\lambda) = B(\lambda) + A(\lambda)\, X(t - \tau(\lambda))
\ .
\end{equation}
Here $B(\lambda)$ is the background spectrum, $A(\lambda)$ is the rms amplitude spectrum of the light-curve variations, and $\tau(\lambda)$ is the time delay spectrum of the reverberations.
With $N$ light curves at wavelengths $\lambda_i$ there are $4\,N$ parameters: 
$\Delta$ for the ROA flexibility, the $N$ mean and rms fluxes $B_i$ and $A_{i}$,
the $N-1$ lags $\tau_{i}$, relative to the designated driving light curve with $\tau=0$,
and finally $N$ extra variances $\sigma^2_i$, 
added in quadrature with the nominal uncertainties, to allow for systematic errors and reduce the influence of light-curves that fail to conform to the ROA model. 
\texttt{PyROA} samples posterior distributions for these parameters using the MCMC sampler \texttt{emcee} \citep{2013PASP..125..306F} with uniform priors on the parameters and the a Bayesian Information Criterion loss function to avoid overfitting.

We ran \texttt{PyROA} on the LCO light-curves using the default Dirac delta transfer function. 
We found it necessary to impose a strong degree of smoothing on the model light-curves by placing a uniform prior, $20 \leq \Delta \leq 30$ days, to fix a wide Gaussian window. 
This was needed to prevent over-fitting of the light-curves which was a problem due to the limited amplitude of the fitted signal relative to the noise in the data. 
This is a consequence of the large size of 3C\,273 which gives it a long expected variability timescale in the accretion disc, and thus less variability over typical observing periods. 
This results in a small amplitude and relatively featureless form of the LCO light-curves in Fig.\,\ref{fig:lc}. 
We set uniform priors on the lags $0<\tau_{i}<100$\,days to include up to $4 \times$ the predicted lags so as to accommodate the potential appearance of the `accretion disc size problem'. For each light-curve the Markov chains were initialised at the predicted lag value, corresponding to the fourth column in Table~\ref{table:lags}, so as to achieve faster convergence. There was no significant difference between these lag estimates and those obtained using the default initialisation value of 0 days. The following analysis was run with 50 chains each with $5\times10^4$ steps in the MCMC sampler. This is significantly more iterations than was used for \texttt{Javelin} due to \texttt{PyROA}'s much faster running time. The optimised parameter values are determined as medians of the MCMC samples with upper and lower uncertainties at the $84^{\mbox{th}}$ and $16^{\mbox{th}}$ percentiles respectively. 

The \texttt{PyROA} and \texttt{Javelin} lags for each year of monitoring are collected in Table~\ref{table:A1}. 
The \texttt{PyROA} lags generally have smaller uncertainties but show relatively poor consistency across the 4 years compared with the \texttt{Javelin} lags. 
This suggests that \texttt{Javelin} may be better suited to fitting the low-variability light-curves (see Table~\ref{table:fvar}), perhaps due to the greater flexibility afforded to it through the use of a Gaussian Process.

\subsection{Variability amplitudes and season selection} \label{Fvar}

Following the approach of \citet{Vaughn2003}, we define the variability of each light-curve using the fractional rms amplitude:

\begin{eqnarray}
F_{\rm var} = \frac{\sqrt{S^{2} - \bar{\sigma}^{2}}}
{\bar{F}
}\ ,
\end{eqnarray}

\noindent
where $\bar{F}$ is the sample mean, $S^2$ is the sample variance, and $\bar{\sigma}^2$ is the mean squared uncertainty of the light-curve flux data. The $F_{\rm var}$ values for light-curves in 6 bands across 4 observing seasons are given in Table~\ref{table:fvar}, quantifying the relatively small variability amplitudes of just a few percent.

\begin{table}
\begin{center}
\caption{Fractional rms amplitude of variations, $F_{\rm var}$, for each photometric filter across all four observing seasons for the light-curves shown in Fig.\,\ref{fig:lc}.}
\label{table:fvar}
\begin{tabular}{||c|c|c|c|c||} 
 \hline
 Filter & \multicolumn{4}{c|}{$F_{\rm var}$ (\%)} \\
 \cline{2-5} \\[-2ex]
 & \emph{Year 1} & \emph{Year 2} & \emph{Year 3} & \emph{Year 4}
 \\ \hline
 $B$ & $2.5$ & $2.3$ & $4.6$ & $3.7$
 \\ 
 $g$ & $2.5$ & $2.5$ & $4.0$ & $2.5$
 \\ 
 $V$ & $1.8$ & $2.1$ & $3.1$ & $2.0$
 \\ 
 $r$ & $1.7$ & $2.5$ & $3.1$ & $2.2$
 \\ 
 $i$ & $1.5$ & $2.0$ & $3.0$ & $2.4$
 \\ 
 $z_s$ & $2.0$ & $2.7$ & $3.1$ & $3.0$
 \\ \hline
\end{tabular}
\end{center}
\end{table}

The third observing season exhibits the largest variability amplitude $F_{\rm var}$ and the clearest features for the RM algorithms to fit. We therefore expect Year\,3 to deliver the most reliable RM results. This expectation is realised for the \texttt{Javelin} lags, which are consistent within their $1\sigma$ uncertainties across all seasons and show the smallest uncertainty for the Year\,3 lags. The \texttt{PyROA} results show far less consistency across seasons but have the best agreement with the \texttt{Javelin} lags in Year\,3 (consistent within $1\sigma$ uncertainties). In contrast, the \texttt{PyROA} lags for Years 1 and 2 overlap with their neighbouring filters within the uncertainty limits. This suggests insufficient variability in these data-sets for the fitting software to clearly differentiate the inter-band lags. Similarly, we attribute the unrealistically large lag estimates obtained by the \texttt{PyROA} Year\,4 fits to a lack of sufficient light-curve variability. For these reasons we opted to focus on the Year\,3 data-set to carry out the remaining analysis. It should also be stated that the variability amplitude demonstrated in our light-curve data falls below the $\sim 10\%$ threshold recommended for adequate RM analysis as estimated by \citet{1997ApJS..110....9R} and is in fact the same order of magnitude as (but still greater than) the typical percentage error on the data points. This difficulty in measuring lags is an inescapable consequence of the large size of 3C\,273 which makes it likely to exhibit less variability over typical observing periods compared to other AGN at similar redshifts. Nevertheless, we were able to obtain reasonable results despite the low variability amplitude ($\sim 3-5\%$), demonstrating that RM on large discs in AGN is feasible when the cadence and data quality are sufficient \citep[see also, e.g.,][]{Fausnaugh2018, 2020MNRAS.497.2910H}.

\begin{figure*}
    \centering 
    \includegraphics[width=0.8\linewidth]{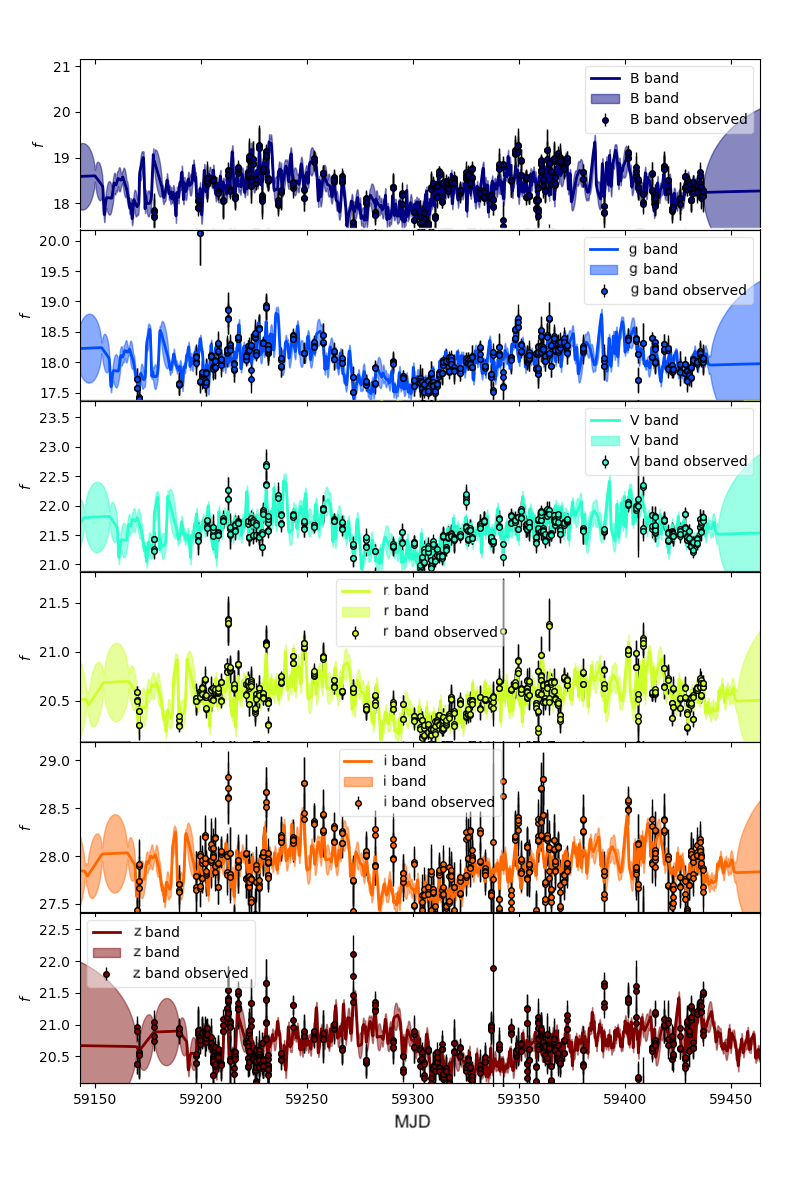}
    \caption{\texttt{Javelin} fits to the detrended Year\,3 light-curves observed for 3C\,273. Fluxes are in mJy and time is represented as modified Julian date.}
    \label{fig:javelin_fit}
\end{figure*}

\begin{figure*}
    \centering 
    \includegraphics[width=0.8\linewidth]{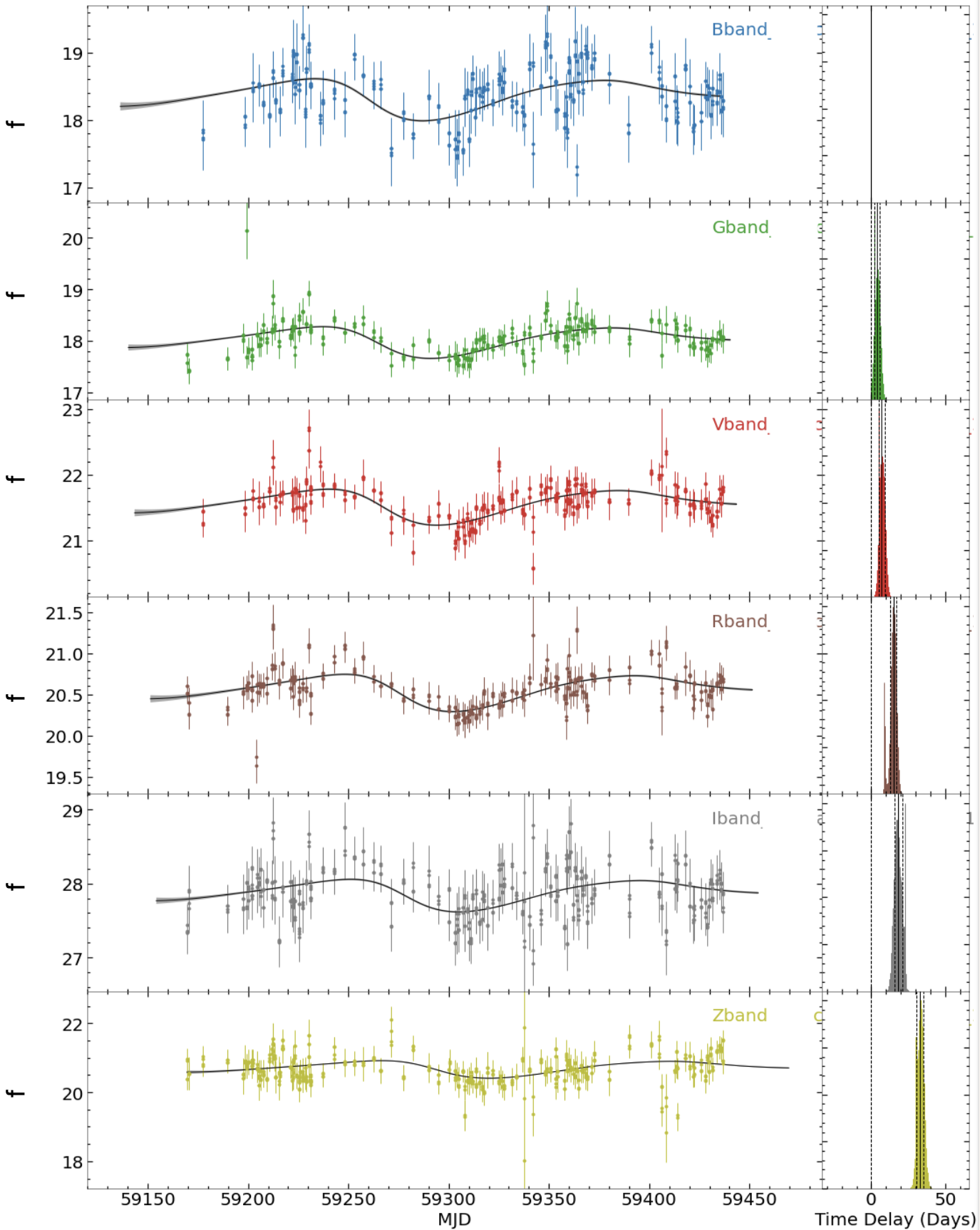}
    \caption{\texttt{PyROA} fits of the model $F(\lambda,t)=B(\lambda) + A(\lambda)\, X(t-\tau(\lambda))$ to the detrended Year\,3 light-curves observed for 3C\,273.  Fluxes are in mJy and time is represented as modified Julian date. 
    The light-curve shape $X(t)$ is a running optimal average over all the data, with a Gaussian width $\Delta=30$\, days.
    The right column of panels show histograms of MCMC samples of the lag parameters $\tau(\lambda)$.
    \label{fig:pyroa_fit}}
\end{figure*}

Tables \ref{table:lags} and \ref{table:trend} summarise the lag estimates obtained by the RM algorithms \texttt{PyROA} and \texttt{Javelin} for the Year\,3 data-set. Also shown are the sets of differential lag predictions obtained with the two treatments of $X$ as explained in Section~\ref{theory}. Fig.\,\ref{fig:javelin_fit} and Fig.\,\ref{fig:pyroa_fit} show the corresponding fits obtained by \texttt{Javelin} and \texttt{PyROA} respectively. The parameter covariance plots obtained by \texttt{PyROA} are shown in the Appendix Fig.\,\ref{fig:pyroa_cov} and the posterior distributions obtained from \texttt{Javelin} are shown in the Appendix Fig.\,\ref{fig:jav_par}. In general, \texttt{Javelin} produced wider posterior distributions which we attribute to the use of a variable-width top-hat transfer function in the fitting procedure compared to the Dirac delta function used by \texttt{PyROA}. This means \texttt{PyROA} applied the same blurring across all light-curves whilst \texttt{Javelin} prescribed separate blurring for each light-curve, meaning that it may produce less constrained lag estimates. We note that the Javelin DRW model seems to overfit the variability, with lots of very fast variability in the model light-curve (Fig.\,\ref{fig:javelin_fit}).
Our \texttt{PyROA} fit, on the other hand, is quite the opposite, and has a very smooth, slowly varying model light-curve (Fig.\,\ref{fig:pyroa_fit}), due to restricting $20<\Delta<30$\,days. Despite this, both 
\texttt{Javelin} and \texttt{PyROA} arrive at consistent lags in year 3, which suggests that assumptions about the underlying variability do not have a large impact on the measured lags in this source.

\begin{table}
\begin{center}
\caption{Differential lag estimates (in observed frame) with respect to the $B$ band obtained by \texttt{Javelin} and \texttt{PyROA} fits to Year\,3 light-curves. The \texttt{PyROA} fits include a $u$-band light-curve and a copy of the $B$-band light-curve, to compute lags and uncertainties for $u$ and $B$. Also included are (observed frame) lag predictions $\tau \propto (X\, \lambda)^{4/3}$ for $X=4.96$ and $X=2.49$ and an estimate of the rest-frame reference lag required to bring the observed lags in agreement with theoretical predictions.
\label{table:lags}
}
\renewcommand{\arraystretch}{1.5}
\begin{tabular}{||c | c | c | r | r | c ||} 
 \hline
 Filter & \multicolumn{2}{c|}{Lag Estimates (days)} & \multicolumn{2}{c|}{Lag Predictions (days)} & $\tau_0$
 \\ \cline{2-5} 
 & \texttt{PyROA} & \texttt{Javelin} & \emph{$X=4.96$} & \emph{$X=2.49$} & (days)
 \\ \hline
$u$ & $-1.9^{+1.9}_{-1.8}$ 
    &  
& -2.52 & -1.00 & 6.8
\\ 
$B$ & $0.0\pm1.9$
    & 0.00 
    & 0.00 &  0.00 &
\\ 
 $g$ & $4.0^{+1.8}_{-2.0}$ 
   & $3.2^{+0.4}_{-0.6}$
& 1.32 & 0.53 & 24.5
 \\
 $V$ & $7.2\pm1.8$ 
    & $6.9^{+1.5}_{-3.3}$ 
& 3.58 & 1.43 & 17.6
 \\
 $r$ & $15.0^{+1.7}_{-2.1}$ 
    & $16.0^{+3.3}_{-9.5}$ 
& 6.26 & 2.50 & 22.2
 \\
$i$ & $17.9^{+2.9}_{-2.4}$ 
    & $15.3^{+15.3}_{-7.6}$ 
& 11.16 & 4.45 & 13.3
\\
 $z$ & $33.2^{+2.1}_{-2.4}$ & $39.6^{+3.1}_{-6.4}$ 
& 15.66 & 6.25 & 20.8
 \\ \hline
\end{tabular}
\end{center}
\end{table}

\begin{table}
\begin{center}
\caption{Differential lag estimates (in observed frame) obtained by \texttt{Javelin} and \texttt{PyROA} with respect to the $B$ band for Year\,3 both with and without linear detrending.
\label{table:trend}
}
\renewcommand{\arraystretch}{1.5}
\begin{tabular}{||c|c|c|c|c||} 
 \hline
 Filter & \multicolumn{2}{c|}{\texttt{PyROA} Lag (days)} & \multicolumn{2}{c|}{\texttt{Javelin} Lag (days)} \\
 \cline{2-5} 
 & \emph{Not Detrended} & \emph{Detrended} & \emph{Not Detrended} & \emph{Detrended}
 \\

 $g$ & $2.6^{+1.8}_{-1.3}$ 
    & $4.0^{+1.8}_{-2.0}$ 
    & $4.3^{+0.9}_{-1.0}$ 
    & $3.16^{+0.40}_{-0.60}$
\\ 
 $V$ & $5.1^{+1.9}_{-1.8}$ 
    & $7.2\pm1.8$ 
    & $11.1^{+1.3}_{-1.8}$ 
    & $6.9^{+1.5}_{-3.3}$
 \\ 
 $r$ & $14.1^{+1.8}_{2.1}$ 
    & $15.0^{+1.7}_{-2.1}$ 
    & $11.8^{+7.6}_{-4.0}$ 
    & $16.0^{+3.3}_{-9.5}$ 
 \\ 
 $i$ & $16.1^{+2.2}_{-2.4}$ 
    & $17.9^{+2.9}_{-2.4}$ 
    & $38.0^{+3.5}_{-13.4}$ 
    & $15.3^{+15.4}_{-7.6}$ 
 \\
 $z$ & $29.6^{+2.5}_{-2.9}$ 
    & $33.2^{+2.1}_{-2.4}$ 
    & $41.6^{+6.9}_{-5.4}$ 
    & $39.6^{+3.1}_{-6.4}$ 
 \\ 
 \hline
\end{tabular}
\end{center}
\end{table}

\subsection{Flux-Flux analysis and spectral energy distributions} \label{FVG and jet}

\begin{figure}
    \centering 
    \includegraphics[width=\columnwidth]{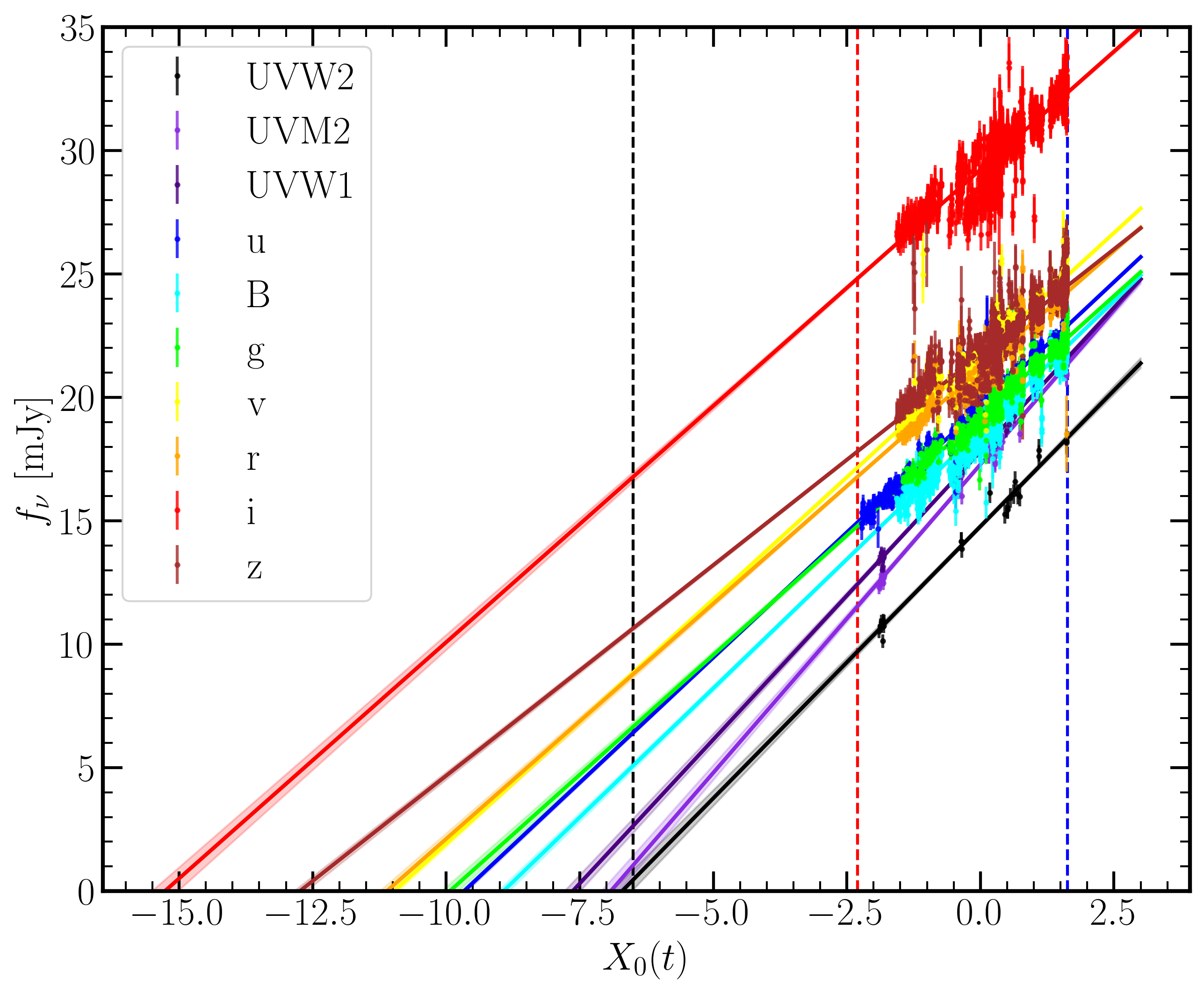}
    \caption{Flux-flux analysis of 3C\,273 based on correlated variations in 3-band UV fluxes from
    \textit{Swift} and 7-band optical fluxes from LCO. Fluxes at different brightness levels are plotted against the model driving light-curve $X_{0}(t)$ (running optimal average fitted with \texttt{PyROA}).  The fitted lines, with shaded 1$\sigma$ uncertainty envelopes, extrapolate the observed fluxes to fainter levels. Vertical lines show the values of the driving light-curve used to evaluate the galaxy contribution (black) and the variable AGN faint (red) and bright (blue) flux. The resulting SEDs are shown in Fig.\,\ref{fig:SED}.
    \label{fig:fluxflux}}
\end{figure}

The assumption made when performing accretion disc RM is that the light-curve variability is dominated by the accretion disc, which should exhibit a characteristic $F_\nu\propto\nu^{1/3}$ spectrum. To test this assumption we use the flux variation gradient (FVG) method \citep{1992MNRAS.257..659W,1997MNRAS.292..273W,2007MNRAS.380..669C} applied to the multi-year LCO light-curves to decompose the spectrum, given by the multi-band photometry, into a variable component (accretion disc) and a non-varying component (host galaxy). Following the approach of \citet{2020MNRAS.497.2910H} and \citet{JVHS2020}, we fit the light-curve flux data at wavelength $\lambda$ with the linear model:
\begin{eqnarray}
\label{FVG}
    F_{\nu}(\lambda,t) = \bar{F}_{\nu}(\lambda) + \Delta F_{\nu}(\lambda)\,X_0\left(t- \tau(\lambda)\right) 
\ .
\end{eqnarray}
\noindent
Here $\bar{F}_{\nu}(\lambda)$ is the mean spectrum, $\Delta F_{\nu}(\lambda)$ is the rms spectrum (i.e., the variable component of the flux) and the light-curve shape defined by $X_0(t)$ is normalised to zero mean and unit variance, and can be time-shifted by $\tau(\lambda)$ to account for the inter-band lags.
We perform the FVG analysis with the `fluxflux' function in \texttt{PyROA}, thus using
Eqn.~(\ref{eq:fbax}) with
the fitted running optimal average, $X(t)$ in Eq.\,(\ref{eq:roa}), for the driving light-curve, with a Gaussian width $\Delta=30$\,days.
The resulting fits from Eq.\,(\ref{FVG}) are shown in Fig~\ref{fig:fluxflux} which plots the observed fluxes in each passband against the dimensionless brightness level, $X_{0}(t)=X(t-\tau(\lambda))$.

As shown in Fig\,\ref{fig:fluxflux},
the relationship between $X_{0}(t)$ and the lagged light-curves is well described by the assumed linear model. The slope gives the rms spectrum $\Delta F_{\nu}(\lambda)$ and the intercept gives the mean spectrum ${\bar{F}_{\nu}}(\lambda)$.
We can also use the model to extrapolate to fainter disc brightness levels.
Extrapolating the shortest-wavelength linear model to fainter flux levels until it crosses the x-axis (i.e., has zero flux) gives an estimate for the lower limit on the non-varying component of the flux. The full spectrum for the non-varying component ($F_{\text{gal}}$) is thus obtained from a vertical cut located at this limit. 
Note that this relatively red SED includes starlight from the host galaxy.
Taking vertical cuts corresponding to the maximum  and minimum  flux values for the light-curves, and subtracting $F_{\text{gal}}$, then gives our SED estimate for the variable (disc) component in its bright state ($F_{\text{bright}}$) and faint state ($F_{\text{faint}}$), respectively. 
Note that the slopes in Fig.\,\ref{fig:fluxflux} are roughly the same at all wavelengths, indicating that the variable SED is approximately flat in $F_\nu$.

\begin{figure}
    \centering 
    \includegraphics[width=\columnwidth]{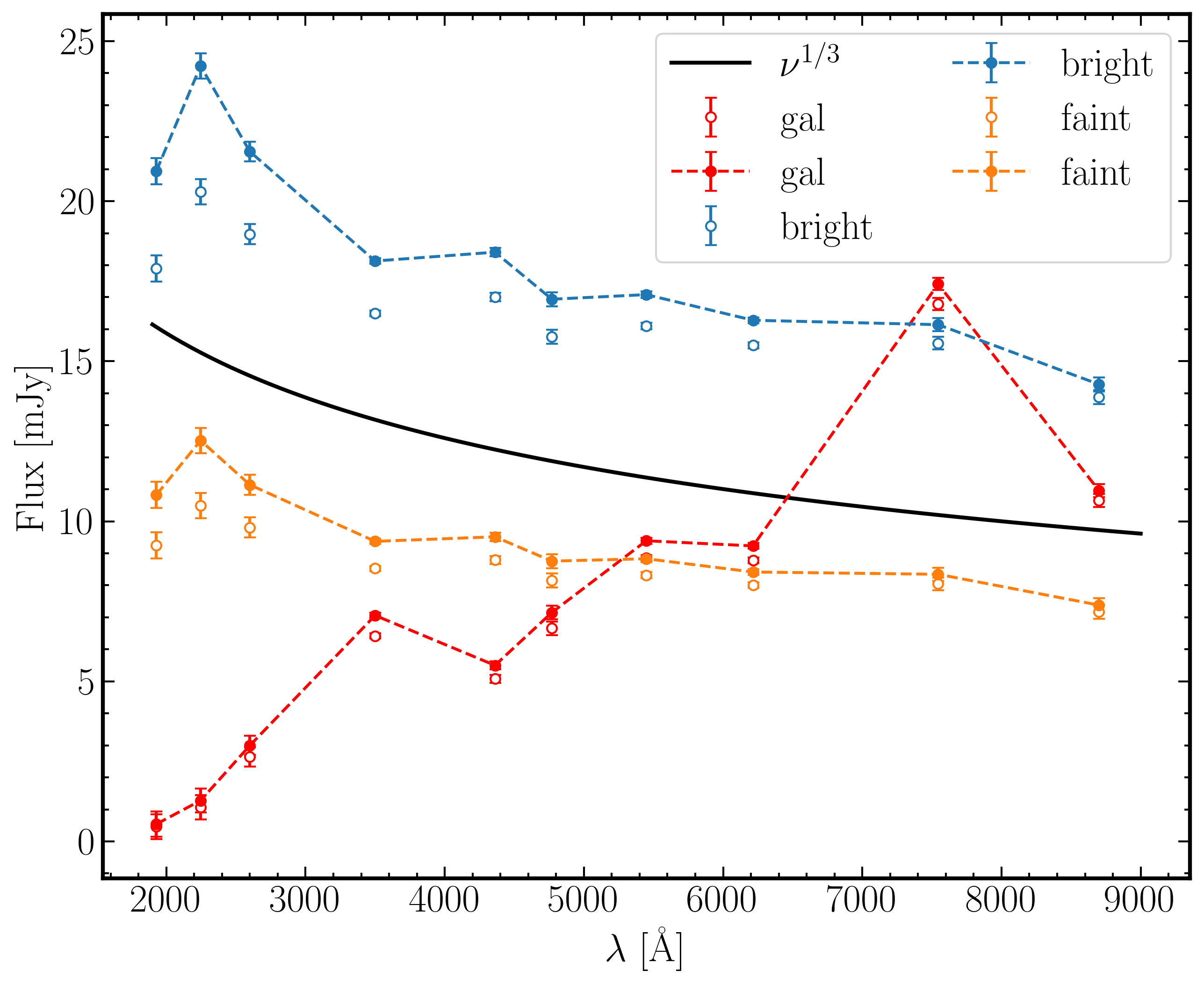}
    \caption{SEDs for the host galaxy (red) and the AGN component in faint (orange) and bright (blue) state inferred from the flux-flux analysis shown in Fig.\,\ref{fig:fluxflux}. All SEDs are de-reddened with a Milky Way reddening law where $E(B-V)=0.021$ \citep{Schlegel1998}. Observed and de-reddened fluxes are plotted as open and filled symbols, respectively. As a reference, the $\nu^{1/3}$ power-law is plotted as black solid curve. The shape of both de-reddened bright and faint AGN contribution to the SED agrees with the reference $\nu^{1/3}$ power-law, suggesting that the variability in 3C\,273 is dominated by the accretion disc. The excess in the galaxy $i$-band flux is due to the contamination of the H$\alpha$ broad emission line
    (see Fig.\,\ref{fig:mike}). 
    \label{fig:SED}}
\end{figure}

For our flux-flux analysis, we use not only the LCO light-curves in 7 optical bands but also archival data in 3 UV bands (W1, M2, W2) from the {\it Swift} satellite. 
The {\it Swift} data are too coarsely sampled in time for UV lag measurements (see Fig.~\ref{fig:uv_lc}), but they do provide useful UV flux measurements at several brightness levels that allow the flux-flux analysis to extend SED estimates into the UV. We note that 3C\,273 has a very low extinction from dust in our Galaxy with a measured value of $E(B-V)=0.021$ \citep{Schlegel1998}. 
Our final dust-corrected SEDs are 
collected in Table\,\ref{tab:fluxflux} and shown in Fig.~\ref{fig:SED}. We note that the recovered optical fluxes for the bright state of the AGN are very similar to the MIKE spectrum shown in Fig.~\ref{fig:mike}. We attribute the non-variable component mainly to the host galaxy, although diffuse continuum emission due to free-free and bound-free transitions in the BLR may also contribute to the non-variable spectrum \citep{JVHS2020}. The derived UV-optical SED for the variable (disc) component of the light-curves is close to the standard thin-disc SED, $F_\nu\propto\nu^{1/3}$, suggesting that the variability in our light-curves is indeed dominated by the accretion disc.

\begin{table}
\begin{center}
\caption{Extinction corrected host galaxy and AGN flux from flux-flux analysis for \textit{Swift} and LCO light-curves. The rms variability amplitude, $\Delta F_\nu$, is measured from the slope of the flux-flux plot analysis in Fig.\,\ref{fig:fluxflux}. The extinction correction assumes a Milky Way reddening law with $E(B-V)=0.021$ \citep{Schlegel1998}. 
\label{tab:fluxflux}}
\renewcommand{\arraystretch}{1.2}
\begin{tabular}{||c | c | c | c | c ||} 
 \hline
 Filter & {$F_{\rm gal}$ (mJy)} & {$F^{\rm faint}_{\rm AGN}$ (mJy)} & {$F^{\rm bright}_{\rm AGN}$ (mJy)} &
 {$\Delta F_\nu$ (mJy)}
 \\ \hline\hline
$UVW2$ & $0.53\pm 0.44$ & $10.83\pm 0.54$ & $20.96\pm 0.55$ & $2.20\pm 0.06$\\

$UVM2$ & $1.25\pm 0.46$ & $12.53\pm 0.57$ & $24.24\pm 0.58$ & $2.50\pm 0.06$\\

$UVM1$ & $2.99\pm 0.35$ & $11.15\pm 0.41$ & $21.56\pm 0.42$ & $2.33\pm 0.05$\\
\hline
$u$ & $7.06\pm 0.06$ & $9.38\pm 0.07$ & $18.13\pm 0.07$ & $2.03\pm 0.01$\\ 

$B$ & $5.50\pm 0.14$ & $9.52\pm 0.16$ & $18.40\pm 0.17$ & $2.09\pm 0.02$ \\ 
 
$g$ & $7.15\pm 0.11$ & $8.76\pm 0.12$ & $16.94\pm 0.13$ & $1.94\pm 0.02$ \\

$V$ & $9.39\pm 0.70$ & $8.83\pm 0.08$ & $17.08\pm 0.08$ & $1.98\pm 0.01$\\

$r$ & $9.23\pm 0.10$ & $8.42\pm 0.11$ & $16.28\pm 0.11$ & $1.91\pm 0.01$\\
 
$i$ & $17.41\pm 0.13$ & $8.35\pm 0.14$ & $16.14\pm 0.15$ & $1.91\pm 0.02$\\

 $z$ & $10.96\pm 0.15$ & $7.38\pm 0.16$ & $14.28\pm 0.16$ & $1.71\pm 0.02$\\ 
 \hline
\end{tabular}
\end{center}
\end{table}

\subsection{Structure function analysis} \label{structure}

\begin{figure}
    \centering 
    \includegraphics[width=\columnwidth]{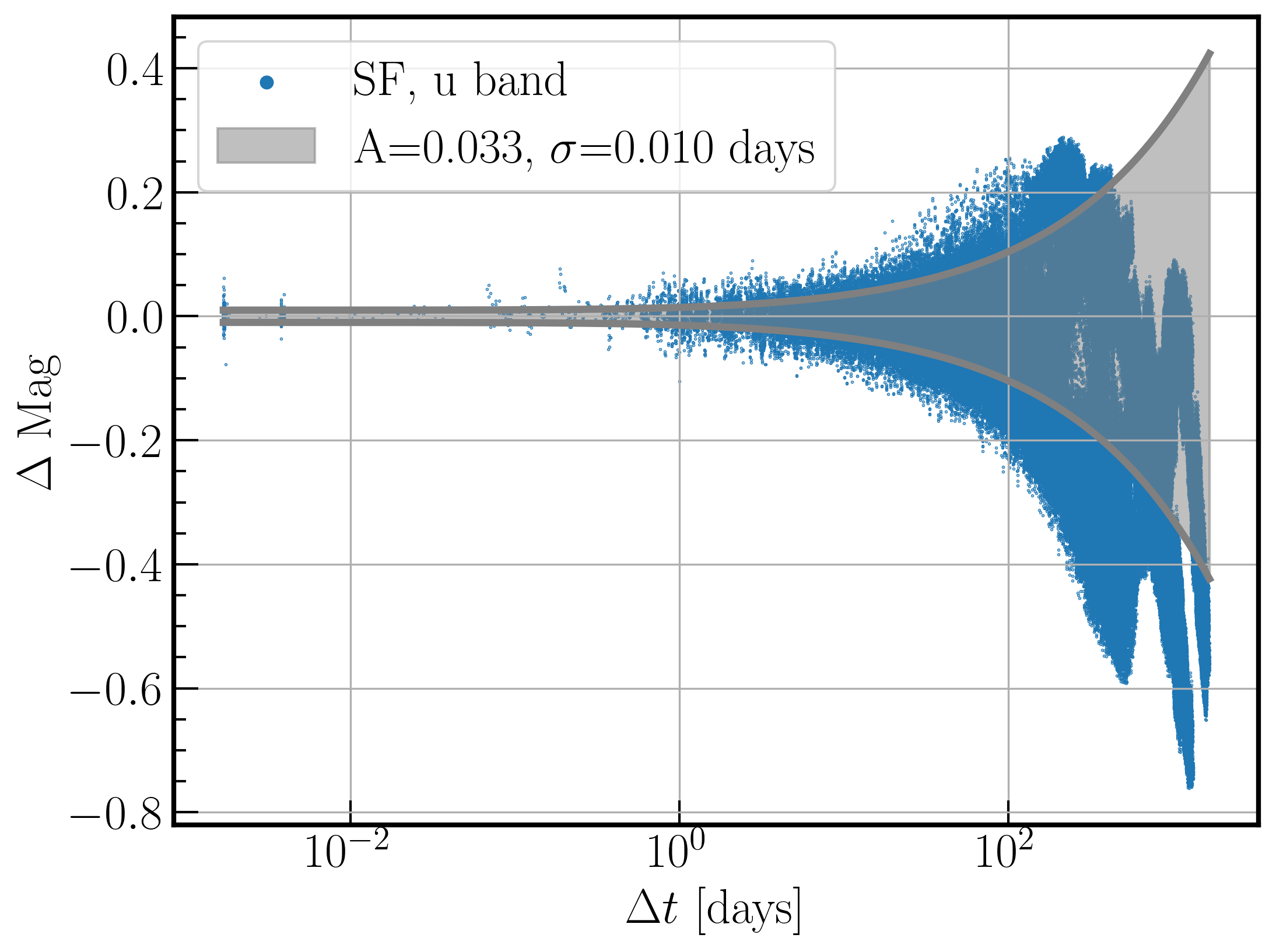}
    \caption{Structure function (SF) analysis of the $u$-band light-curve. Each blue dot corresponds to the change in the magnitude $\Delta m_{ij}$ for a pair of data points separated in (rest-frame) time by $\Delta t = (t_j - t_i)/(1+z)$ with $t_j > t_i$. The grey shaded area shows the fit to the SF variance with a random walk (RW) model, as described in in Eq.\,(\ref{eq:rw}),
    with a 10-day rms amplitude $A=0.033$\,mag and rms noise $\sigma=0.010$\,mag.
    The SF data at $\Delta t \ga 100$ days becomes asymmetric, indicating that the light-curves are too short to provide a good statistical sample of such long-timescale variations. 
    The DRW de-correlation timescale $\tau_\text{d}$ cannot be accurately determined with the current light-curve data.  We therefore report the 10-day RW amplitude $A=0.033$\,mag.
    \label{fig:ubandstructure}}
\end{figure}

To study the variability of 3C\,273 on different timescales, one approach is to use Fourier transforms to acquire power spectra of the light-curves. This method is particularly well-suited for periodic signals, in contrast to the red noise character of AGN light-curves. In addition, the non-uniform time sampling, seasonal gaps, and outliers in the light-curve data complicate the Fourier analysis. We therefore opt for an equivalent method based on the structure function (SF) \citep{Collier2001}. The structure function $S(\Delta t)$ quantifies the changes in brightness that occur after a span of time $\Delta t$.  Given $N$ measured fluxes, $F_i$, at times $t_i$, there are $N(N-1)/2$ flux pairs that give structure function data, the change in flux (or magnitude) over the time span $\Delta t_{ij} \equiv t_j - t_i$. 
The change in magnitude of the AGN disc is
\begin{equation}
    \Delta m_{ij} =
    2.5\,\log_{10}\left( \frac{ F_j - F_\text{gal} }{ F_i - F_\text{gal} }\right) = 
    m(t_i)-m(t_j);~\textup{for}~t_i < t_j
    \ .
\label{eqn:SF}
\end{equation}
\noindent
with corresponding uncertainty $\sigma_{ij} = \sqrt{ \sigma_i^2 + \sigma_j^2}$.
Note that we subtract the galaxy flux $F_\text{gal}$ (see Fig.\,\ref{fig:SED}) from the fluxes $F_i$, so that the magnitude change represents the variation in the AGN disc component, undiluted by the relatively red host galaxy flux.
The blue dots in Fig.\,\ref{fig:ubandstructure} show the magnitude changes from our $u$-band light-curve as a function of rest-frame time separation $(t_j - t_i)/(1+z)$. These scatter around 0 with a variance that increases with timespan $\Delta t$.
For $\Delta t\la1$\,day, $\Delta m_{ij}$ is dominated by the rms measurement uncertainty $\sigma_{ij}$.
For $\Delta t\ga1$\,day
the variance increases due to the intrinsic variations of the AGN.
For $\Delta t\ga100$\,days the distribution becomes asymmetric above and below 0, indicating that the light-curve is not long enough to provide a good statistical sample of intrinsic variations on such long timescales.

AGN variations typically have a red-noise character, with larger amplitudes on longer timescales, that can be characterised with a damped random walk (DRW) model
\citep{Kelly_2009, Macleod_2010, Tie_2018, Burke_2021}.
The DRW structure function model is a Gaussian distribution with zero mean and a variance that increases with the timespan $\Delta t$ as
\begin{equation}
S_\text{DRW}^2(\Delta t)= S_\infty^2\, (\,1 - \exp(- \Delta t/\tau_\text{d})\,)
\ .
\label{eq:drw}
\end{equation}

\noindent
For $\Delta t \la \tau_\text{d}$, the variance increases linearly with $\Delta t$ up to the de-correlation timescale $\tau_\text{d}$ above which it approaches the asymptotic variance $S_\infty^2$.

Our 3C\,273 light curves are too short to measure the DRW de-correlation timescale, providing only a rough lower limit $\tau_\text{d}\ga300$\,days.
 Such a large value of $\tau_\text{d}$ is expected considering the large luminosity of 3C\,273, where the DRW timescale is known to increase with the mass of the SMBH and the AGN continuum luminosity \citep{Macleod_2010, Burke_2021}. 
 3C\,273 also has a moderate redshift, which means a longer period of time coverage is needed due to the cosmic time dilation.

With $\tau_\text{d}$ beyond the range of our 3C\,273 monitoring campaign, we characterise the variations with a random walk (RW) model
\begin{equation} \label{eq:rw}
    S_\text{RW}^2(\Delta t) = A^2\, \left( \Delta t / 10\,\text{d} \right) \ ,
\end{equation}
where $A$ is the rms amplitude of variations over a 10-day timespan.
We use MCMC methods to fit 2 parameters, $A$ and $\sigma$, to the $N(N-1)$ magnitude pairs $\Delta m_{ij}$.
The log-likelihood for this fit is given by
\begin{equation}
-2\,\ln{L(A,\sigma}) 
= \sum_{ij} \left(
\frac{ \Delta m_{ij}^2 }
{ S^2_{ij} + \sigma^2 }
+ \ln{ \left( S_{ij}^2 + \sigma^2 \right) } 
+ \ln{(2\,\pi)}
\right)\ ,
\end{equation}
where $S^2_{ij} = S_\text{RW}^2\left(\Delta t_{ij}\right) = A^2\, \left( \Delta t_{ij}/10\,\text{d}\right)$,
and $\sigma^2$ is the noise variance that dominates for $\Delta t \la 1$\,day.
The grey envelope in Fig.\,\ref{fig:ubandstructure} shows the result for our $u$-band lightcurve data.

\begin{figure}
    \centering 
    \includegraphics[width=\columnwidth]{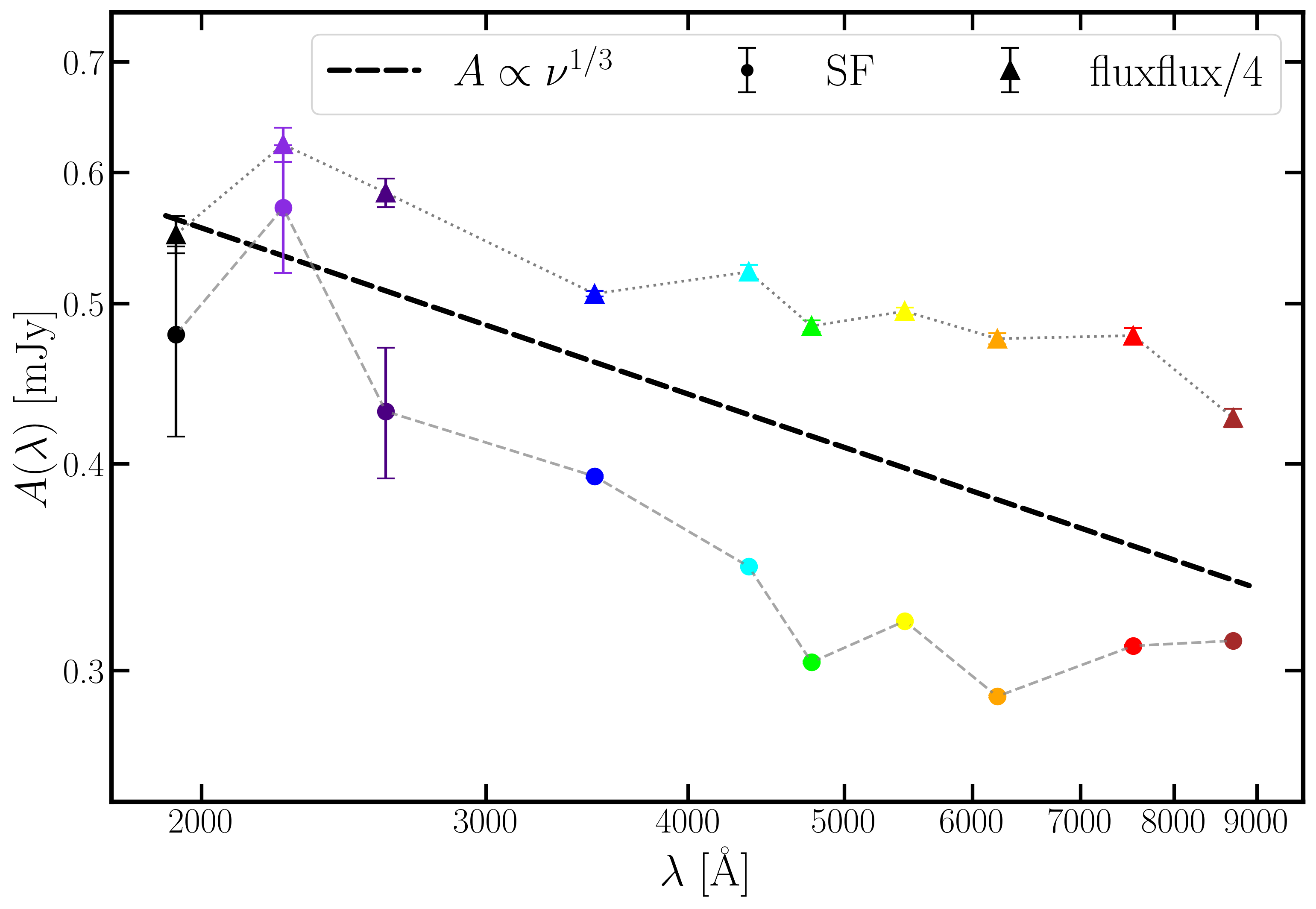}
    \caption{Structure function fitting results for the 10-day rms amplitude $A(\lambda)$ (coloured circles) compared with the rms variability amplitude, $\Delta F_\nu$ (coloured triangles) from the flux-flux analysis (see Table\,\ref{tab:fluxflux}). The amplitude from the flux-flux analysis is multiplied by a factor of 0.25 for better comparison with the RW amplitudes. The expected variation amplitude for the thin disc ($F_\nu\propto \nu^{1/3} \propto \lambda^{-1/3}$ , black dashed line) is plotted as a reference. The wavelength dependency of the variation amplitude is consistent with both analyses.}
    \label{fig:structurespec}
\end{figure}

To investigate the wavelength-dependence of the 10-day RW amplitude $A(\lambda)$, 
we perform the above analysis with each of the UV light-curves from Swift and optical light-curves from LCO. 
The result is plotted in Fig.\,\ref{fig:structurespec} (coloured circles) and compared to our flux-flux analysis results obtained in Section~\ref{FVG and jet} (coloured triangles). Both variability analysis methods yield results roughly consistent with the $F_\nu\propto \nu^{1/3}$ spectrum expected for a geometrically thin, optically thick accretion disc in this wavelength range. 

\section{Discussion} \label{Discussion}

\subsection{The accretion disc size problem} \label{ad size problem}

The `accretion disc size problem' is a discrepancy detected for many, mostly low-luminosity AGN studied so far with accretion disc RM. The inter-band lags imply physical scales for the accretion discs that are a factor of $\sim 2-3$ larger than expected by theoretical predictions \citep[e.g.][]{2016ApJ...821...56F, Jiang_2017, Mudd2018,2018ApJ...857...53C, Edelson_2019}. Given this significant discrepancy, our understanding of how the radius scales with wavelength in the accretion disc may be incomplete or some of the assumptions made in obtaining the lag estimates may be incorrect.

For 3C\,273, our lag estimates obtained with \texttt{PyROA} and \texttt{Javelin} (Table~\ref{table:lags}) are consistent with each other within their uncertainties. 
With the exception of the $u$-band, the lags exceed the predictions made with the standard $X=4.96$ value by a factor of $\sim 2-3$ and those made with $X=2.49$ by a factor of $\sim 4-7$. Therefore, as most other AGN studied so far, 3C\,273 shows the `accretion disc size problem'.
The one exception is the $u$~band, for which the \texttt{PyROA} lag reported in Table~\ref{table:lags} is within the predicted lag range. Often an $u$-band lag excess is seen in other well-studied AGN, which may be attributed to contamination by diffuse Balmer continuum emission from the BLR. Such an effect does not seem to be present in 3C~273. 

To our knowledge, quasars with luminosities close to the Eddington limit have not been studied so far with high-cadence, high-quality disc monitoring such as done here. \citet{Mudd2018} made a first attempt on this front, however, due to the quality of their data, their study yielded only an ambiguous result and was not able to establish a strong preference for or against the `accretion disc size problem'. Later, \citet{Li2021} and \citet{Guo2022} found in their AGN samples, which included also quasars, a trend for less luminous AGN to have larger discrepancies between the observed and predicted lag values, meaning that the `accretion disc size problem' is expected to be less pronounced in luminous quasars. However, \citet{Miller2023} presented an accretion disc RM study of the high-luminosity AGN Mrk~876 accreting at a relatively high fraction of $L_{\rm acc}/L_{\rm Edd} \approx 0.4$ and found lags a factor of $\sim 3$ larger than theoretical predictions, just as previous studies found for the low-luminosity AGN. A similar result was found also by \citet{2020MNRAS.497.2910H} for another source with a similarly high accretion rate, namely the broad-line radio galaxy 3C\,120. On the other hand, \citet{Cackett2023} found in their high-quality data-set on Mrk~817, an AGN accreting at a moderate rate of $L_{\rm acc}/L_{\rm Edd} \approx 0.2$, that the accretion disc lags on short timescales were largely consistent with reverberation from a standard disc. \citet{Kokubo2018} studied the quasar PG~2308$+$098, for which they estimated a relatively low accretion rate of $L_{\rm acc}/L_{\rm Edd} \approx 0.1$, and found that the observed accretion disc lags deviated only moderately from theoretical expectations (by a factor of $\sim 1.2 - 1.8$). Finally, \citet{Donnan2023} performed a high-quality accretion disc RM campaign on the super-Eddington narrow-line Seyfert 1 galaxy PG~1119$+$120 and found only a tentative hint of the `accretion disc size problem'. Clearly, more high-quality accretion disc RM studies of AGN covering a large range in luminosity is required to settle the issue.

\subsection{The suitability of the thin disc model in 3C\,273}

\begin{figure}
    \centering 
    \includegraphics[width=\columnwidth]{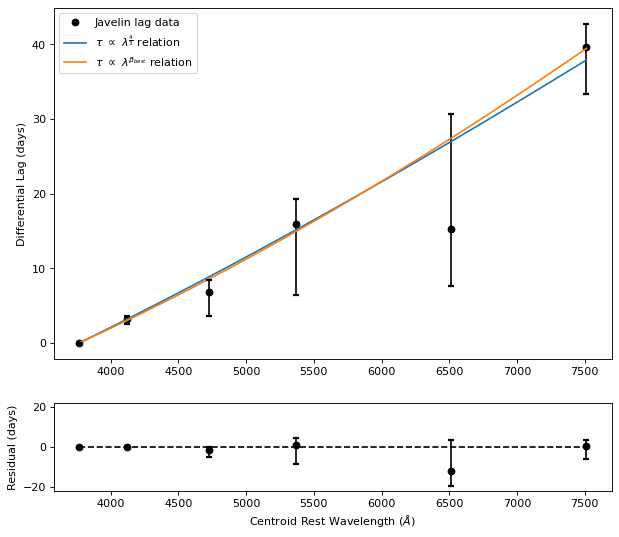}
    \includegraphics[width=\columnwidth]{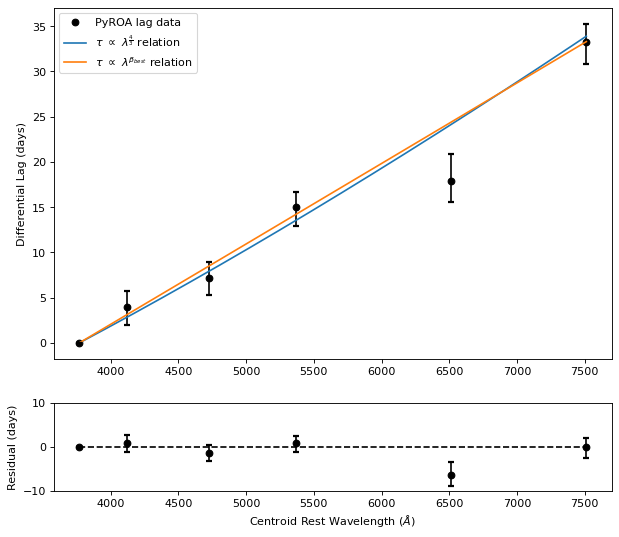}
    \caption{Power-law models, $\tau \propto \lambda^{\beta}$, fit to the RM lag results from Table~\ref{table:lags} as obtained by \texttt{Javelin} (top panel) and \texttt{PyROA} (bottom panel). We omit the $u$-band lag from the fit due to likely Balmer continuum contamination. Shown are fits to the results with the thin disc model exponent $\beta = 4/3$ (blue lines) and fits to the results when $\beta$ is left as a free parameter (orange lines). The residuals in the lower panels are given with respect to the fits when $\beta$ is left to vary.}
    \label{fig:tau-lam}
\end{figure}

The differential lags obtained by \texttt{PyROA} and \texttt{Javelin} (with respect to the $B$ filter) are plotted in Fig.\,\ref{fig:tau-lam} against the rest-frame wavelengths of the LCO light-curves. The physically motivated power-law $\tau \propto \lambda^{\beta}$ has been fitted to the data-sets with the thin disc model exponent $\beta = 4/3$ (blue lines). 
The measured lags are generally consistent with the thin disc model expectation within the $1\sigma$ uncertainties. In addition to the $u$-band lag, which we omit from the fit due to likely Balmer continuum contamination, another possible outlier is the $i$-band lag estimated by \texttt{PyROA}, which has smaller errors than that estimated by \texttt{Javelin}. 
Given that the $i$ band has a strong contamination by the broad H$\alpha$ emission line (see Fig.\,\ref{fig:mike} and Table~\ref{tab:1}) we might expect this lag to be {\it larger}, not smaller.
Therefore, the $i$-band lag estimates appear to be anomalous and we therefore omit them from consideration in the following fits.
  

\begin{table}
\begin{center}
\caption{Estimates of the power-law index $\beta$ in the expected relation $\tau\propto\lambda^{\beta}$ obtained from the RM results from PyROA and Javelin, and from the spectral fit $\nu\,F_\nu\propto\lambda^{-\beta}$ in Fig.\,\ref{fig:spectrum}.}
\label{table:beta}
\resizebox{\columnwidth}{!}{%
\begin{tabular}{||c | c | c | c | c||} 
 \hline
 & Thin disc & \multicolumn{2}{c|}{RM results} & Spectral\\
 \cline{3-4} 
 & model value & \emph{PyROA} & \emph{Javelin} & value $\beta_{\rm spec}$\\
 \hline
 $|\beta|$ & $4/3$ & $1.01\pm0.39$  & $1.59\pm0.29$ & $1.35 \pm 0.13$ \\ 
 \hline
\end{tabular}%
}
\end{center}
\end{table}

Repeating the power-law fits whilst allowing $\beta$ to vary generated the orange curves in Fig.\,\ref{fig:tau-lam}. 
The preferred $\beta$ value fit in each case is stated in Table~\ref{table:beta}. Both the \texttt{PyROA} and \texttt{Javelin} results
fit power laws consistent with the thin disc model within $1\sigma$ uncertainties. The \texttt{PyROA} lag estimates favour a slightly shallower $\beta$ value whilst the \texttt{Javelin} estimates favour a slightly steeper value, but this difference is not significant given the uncertainties. 

\begin{figure}
    \centering 
    \includegraphics[width=\columnwidth]{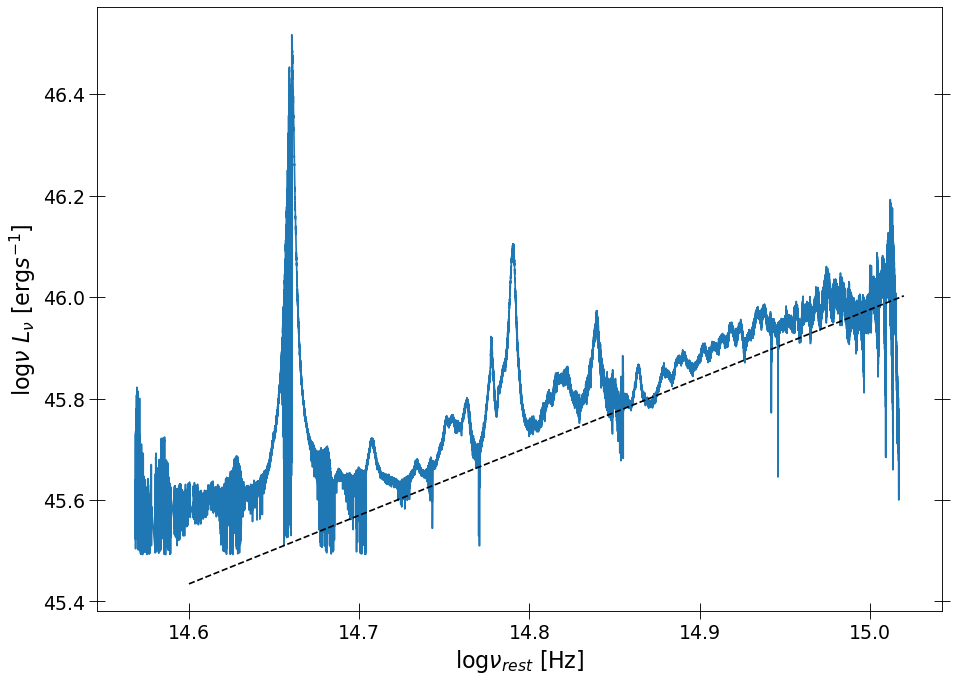}
    \caption{Spectrum of 3C\,273 taken from the MIKE spectrograph plotted in a logarithmic form. A power law is fit which returns a measured power law exponent $\beta = 1.35 \pm 0.13$}
    \label{fig:spectrum}
\end{figure}

The power-law index $\beta$ can be estimated not only from the lag spectrum $\tau\propto\lambda^\beta$, but also independently from the accretion disc's flux spectrum, $\nu\, F_\nu \propto \nu^\beta$.
Fig.\,\ref{fig:spectrum} shows a log-log plot of the contemporaneous MIKE optical spectrum of 3C\,273, plotted in terms of monochromatic luminosity $\nu\,L_\nu$. 
Also shown is a power-law model (black dashed line) that traces the spectral continuum.
The slope on Fig.\,\ref{fig:spectrum}
gives the power-law index $\beta_{\text{spec}} = \text{d}\log{(\nu\,L_{\nu})}/\text{d}\log{\nu}=1.35\pm0.13$, as stated in Table~\ref{table:beta}. 
This spectral exponent estimate is consistent with the theoretical thin disc model expectation ($\beta = 4/3$) and is consistent with the RM results within the uncertainties. Our results can be compared to a previous measurement of the $\beta$ exponent in the accretion disc of 3C\,273 by \citet{Figaredo_2020} which was obtained by fitting power-laws to observed optical \emph{bvrz} bands. They estimated a value of $\beta = 1.34 \pm 0.06$, consistent within the uncertainty limits of our RM results and our spectral value. 
Therefore, our spectral fit and RM results both independently suggest that 3C\,273 is well described by the thin disc model. 


\subsection{Disc model with a steep rim in 3C\,273} \label{rim}

Following \citet{Starkey2023}, in order to address the accretion disc size problem that arose from the thin disc model, we generalise the zero-thickness disc model, $H(r)=0$, to a finite thickness power-law model, with a lamp-post irradiating the disc at $r=0$.
The geometry of the accretion disc is modelled as: 

\begin{equation}
\label{eq_hpowerlaw}
	H(r) =  H_{\rm out} \left(\frac{r}{r_{\rm out}}\right)^k,
\end{equation}

\noindent
where $H_{\rm out}$ is the height of the disc at reference (outer) radius $r_{\rm out}$. 
We model the lamp-post with $\epslp$, which is a dimensionless efficiency factor. The dimensionless energy conversion and reprocessing efficiency is:

\begin{equation}
	\epslp \equiv \frac{ L_{\rm LP} }{ \dot{M}\, c^2},
\end{equation}

\noindent
where $L_{\rm LP}$ is the bolometric luminosity of the two lamp-posts, above and below the black hole, that irradiate the two sides of the disc, and all light rays shining on the disc get absorbed and reprocessed. $\epslp$ can differ from that of the disc:

\begin{equation}
    \eta_{\rm disc}
    \equiv \frac{ L_{\rm disc} } { \dot{M} \, c^2 } \ .
\end{equation}

\noindent
We consider a concave disc where $k > 1$ \citep{Starkey2023}. The irradiative heating term due to the lamp-post is $\epslp\,(r/r_\text{g})\,f(r)$,
where $f(r) = r\, \frac{ \partial }{ \partial r} \left( \frac{ H(r) - H_{\rm LP} }{ r } \right)$
is the covering fraction, and $H_{\rm LP}$ is the height of the lamp-post. Therefore, the effective temperature with the irradiative heating term is
\begin{eqnarray}
    T_{\mathrm{eff}} &=& T_{\mathrm{visc}} + T_{\mathrm{irrad}} \\
    &=& \left( \frac{G \,M \,\dot{M}}{2 \,\pi \,\sigma\, r^3} \right) \left[ \frac{3}{4} \left( 1 - \sqrt{\frac{r_{\mathrm{isco}}}{r}} \right) + \frac{r\,\epslp}{r_{\rm g}}\, f(r) \right]
    \ ,
\end{eqnarray}
where $r_{\rm g} \equiv G\,M\slash c^2$ is the gravitational radius of the SMBH. With $k>1$, the irradiation temperature term increases with $r$ so that $T_{\rm eff}$ decreases with $r$ less steeply than $r^{-3/4}$, the SED becomes redder, and the lags increase.

\begin{figure*}
\centering
	\begin{tabular}{@{}cccccc@{}}
 \multicolumn{1}{l}{}\\
	\includegraphics[width=\columnwidth]{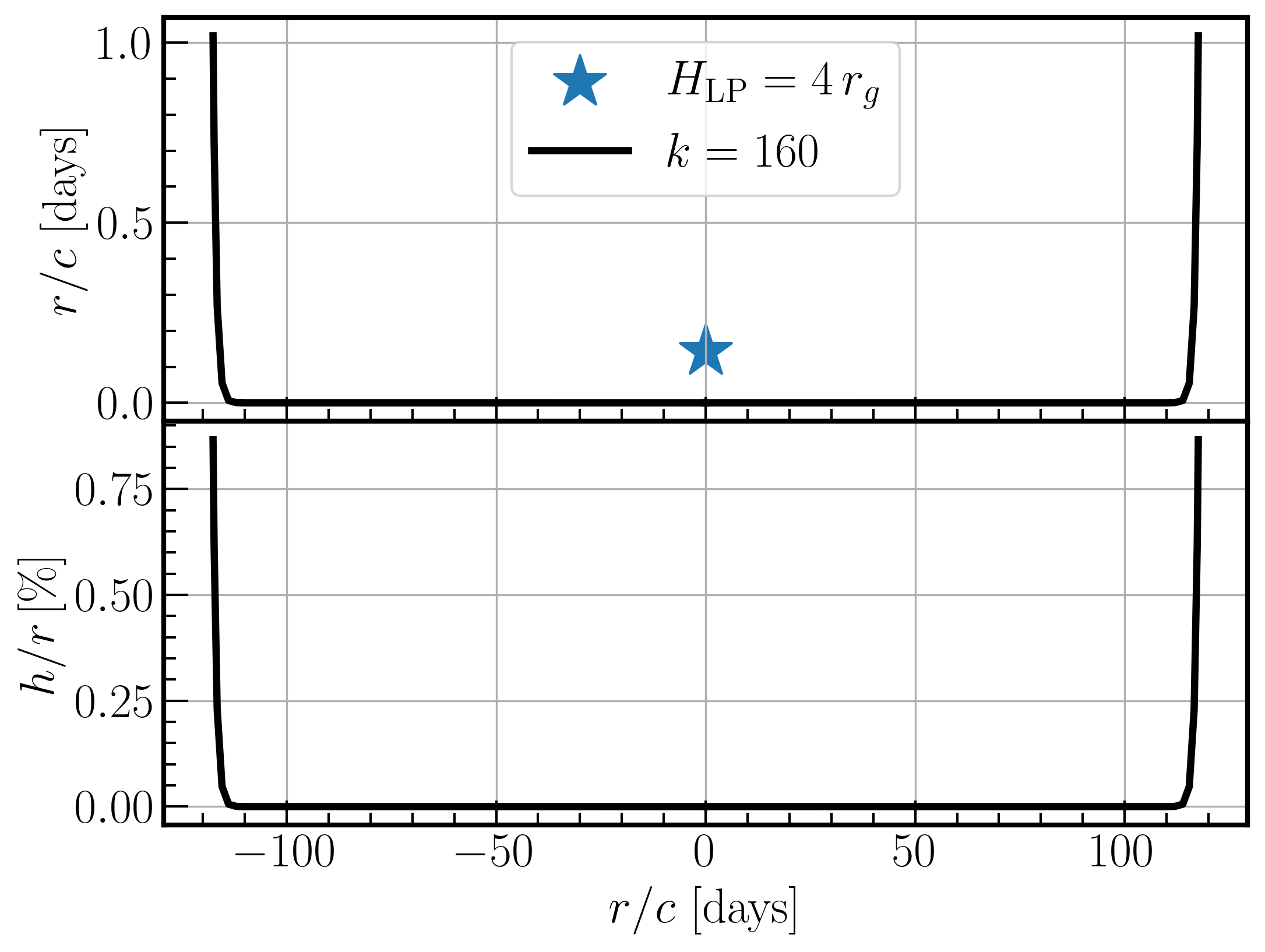}
	\includegraphics[width=\columnwidth]{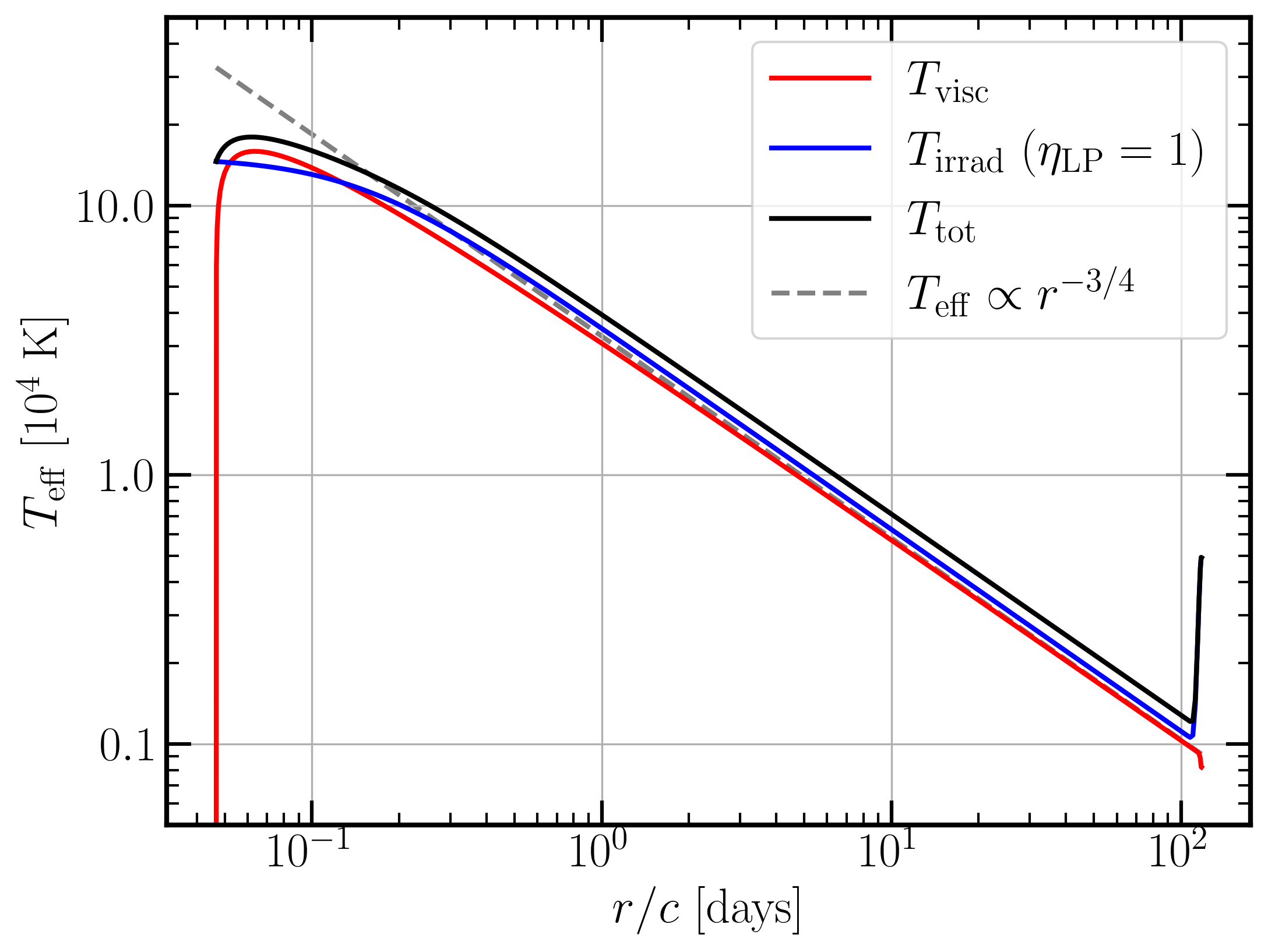}
 \\
  \multicolumn{1}{l}{}\\
	\includegraphics[width=\columnwidth]{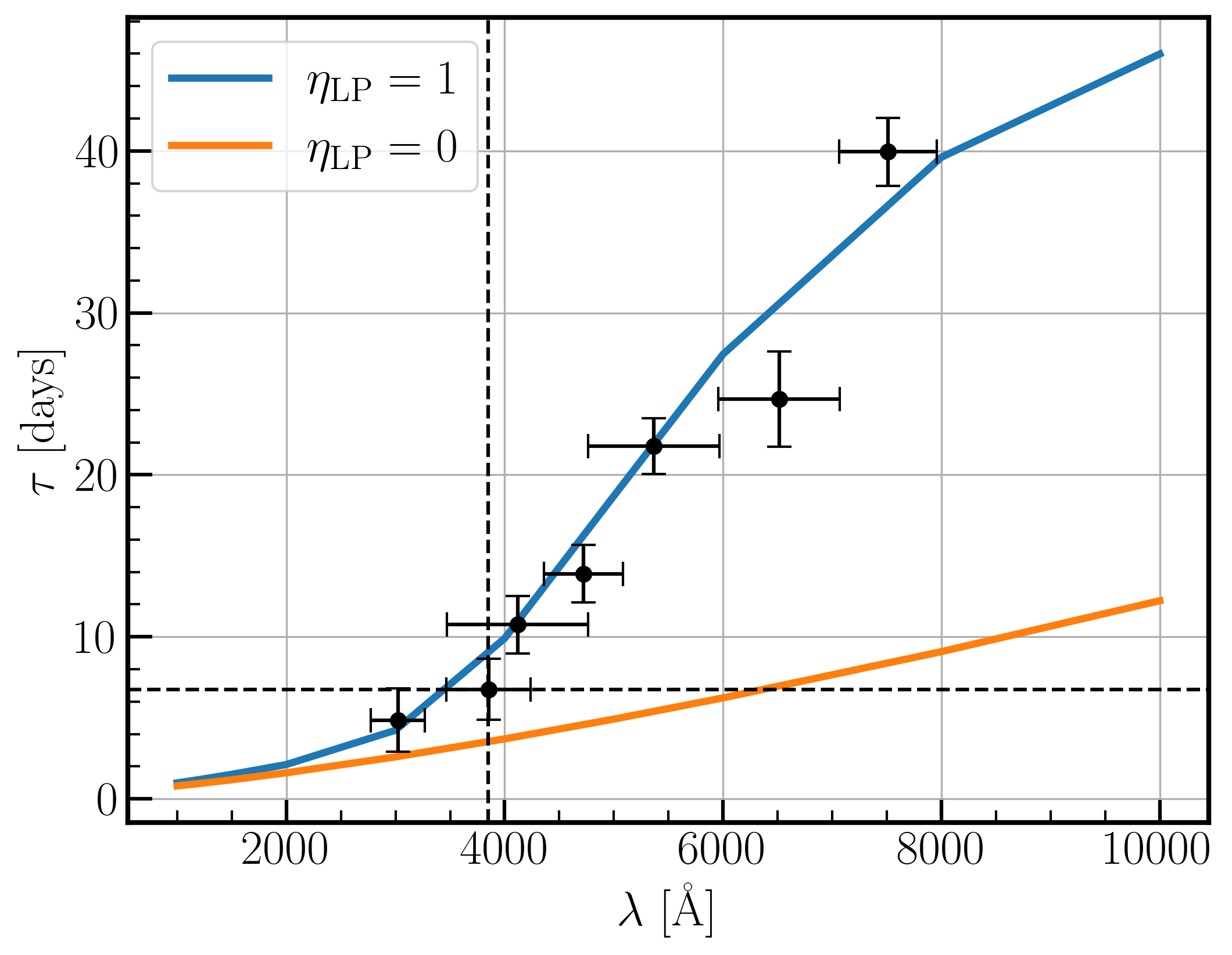}
	\includegraphics[width=\columnwidth]{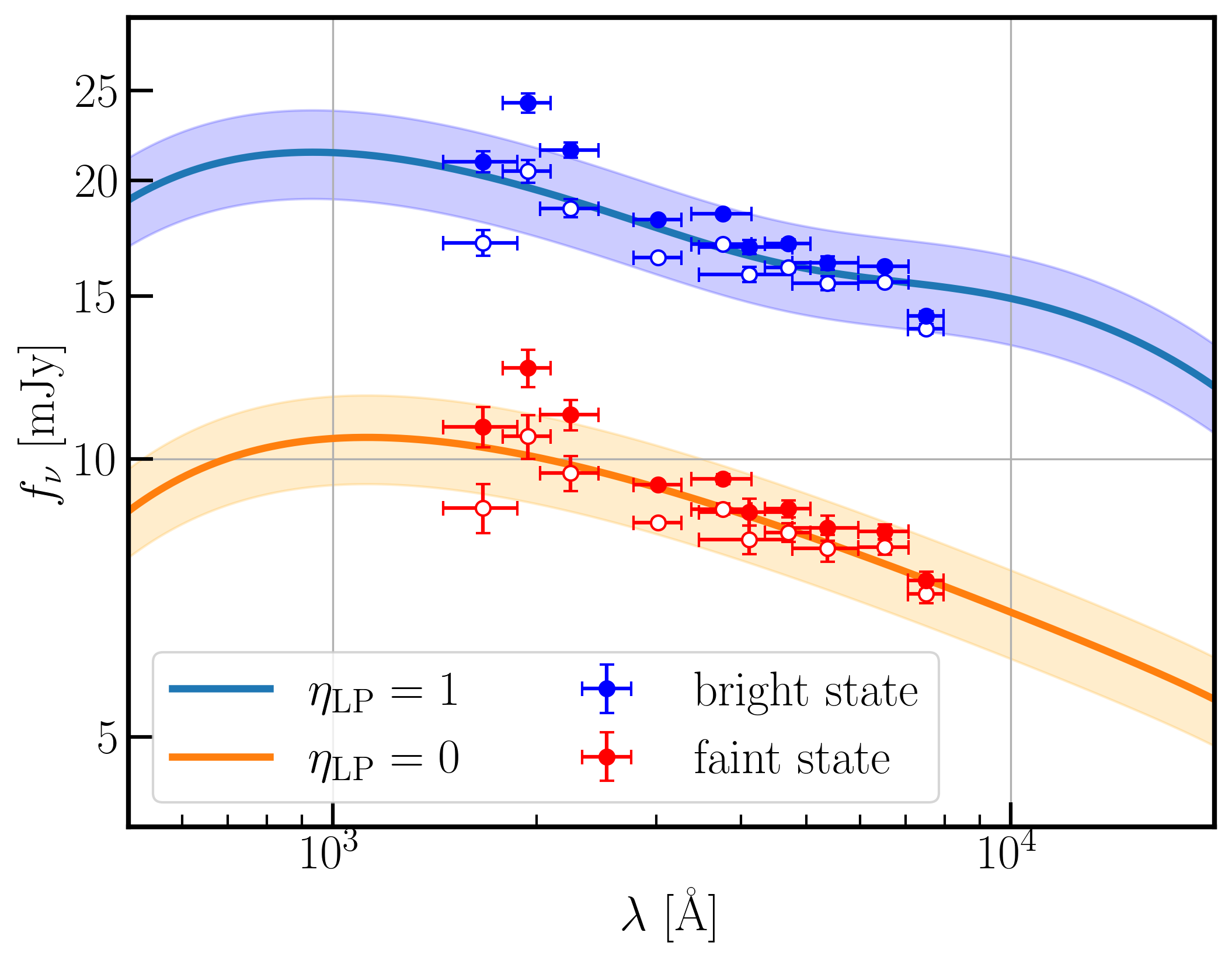}
	\end{tabular}
    \caption{Simultaneous fit to the SED and delay spectrum of 3C\,273. 
    \textbf{\textit{Upper left:}} Geometry of the accretion disc with the fitted parameters. The black curve shows an nearly-flat discs with a relatively steep rim that rises to $H_{\rm out}= H(r_{\rm out}) \approx 1 $~light days ($H/r\approx 0.8\%$) at the outer radius $r_{\rm out}=120$~light days. The disc height parameterized as $H(r)=H_{\rm out}\,\left(r/r_{\rm out}\right)^k$ with a power law index $k = 160$. The blue star marks the location of the lamp-post with height $H_{\rm LP}=4~r_\text{g} \approx 0.2$~light days. The disc inclination is fixed at $i=0^{\circ}$.
    \textbf{\textit{Upper right:}} The effective temperature profile. The red curve shows the viscus heating contribution to the effective temperature, which is equivalent to the temperature profile of the disc when the lamp-post is turned off (i.e. $\epslp=0$ for AGN faint state), with a best-fit accretion rate $\dot{M} \approx 1.6\, \rm M_{\rm \odot}\, yr^{-1}$. For $r\gtrsim 1$ light days, the viscous temperature falls as $T_{\rm eff}\propto r^{-3/4}$ reaching $10^3$~K at $r=r_{\rm out}$ with the lamp-post off. The blue curve shows the temperature contribution due to the lamp-post heating up the accretion disc, with lamp-post radiative efficiency of $\epslp=1$. The temperature rises rapidly to $\sim 5000$~K due to irradiation on the steep inner face of the rim. The black curve shows the total temperature ($T_{\rm tot} = T_{\rm visc} + T_{\rm irrad}$) of the disc when the lamp-post is turned on, and the grey dashed curve with $T_{\rm eff}\propto r^{-3/4}$ is plotted as a reference.
    \textbf{\textit{Lower left:}} The model continuum lag for the faint (orange curve, $\epslp=0$) and bright (blue curve, $\epslp=1$) state. The black error bars show the measured rest frame continuum lag from the Year-3 LCO light curves (See Table\,\ref{table:lags}). The vertical dashed line corresponds to the rest frame wavelength of the reference B-band, and the horizontal dashed line shows the lag for the reference B band, $\tau_{\rm ref} \approx 6.8$ days.
    \textbf{\textit{Lower right:}} The model disc SEDs of the faint-disc data (orange curve) with the lamp-post off, and the bright-disc data (blue curve) with the lamp-post on, with irradiative efficiency $\epslp = 1$. The shaded region shows the fitted uncertainty parameter in the SED modelling ($\sigma_{\rm SED} = 10\%$). The filled red and blue error bars show the extinction corrected AGN SED for the bright and faint state in the rest frame, respectively (See Table\,\ref{tab:fluxflux}). As a reference, we also plot the open error bars showing the SED from flux-flux analysis without extinction correction. The steep-rim disc models achieve a satisfying fit to both the continuum lag and disc SED data. The posterior distribution of the MCMC fit is shown in Fig.\,\ref{fig:bowl_cov}.}
    \label{fig:steep}
\end{figure*}

Following \citet{Starkey2023}, we adopt a three-step approach in fitting simultaneously the SED and the lags, namely, we first adjust the accretion rate and inner radius to match the faint-disc SED with the lamp-post off, then, with the lamp-post on, we fit the bright-disc SED, and, finally, we adjust the $H(r)$ profile to fit the lags. Fig.\,\ref{fig:steep} shows the MCMC fit for a face-on power-law accretion disc model. This best-fit model has $k=160$, which produces a steep rim on the outer edge of the disc at $r_{\rm out}\sim 120$~light-days, for a rest-frame ($B$-band) reference lag of $\tau_{\rm ref} = 6.8$~days. The rim height is $H_{\rm out}=1$~light-day, corresponding to $H/r=0.8\%$. The faint state SED is fitted with $\epslp=0$ and a black hole accretion rate $\dot{M} \approx 1.6\, \rm M_{\rm \odot}\, yr^{-1}$. With $\epslp=1$ for the bright state fit, the declining $T_{\rm eff}\propto r^{-3/4}$ on the flat inner disc is elevated by a factor $\sim1.5$, and rises rapidly to $T_{\rm eff}\sim 5000$~K at the irradiated rim. The mean lag from a vertical rim is: $\left<\tau\right>_{\rm rim}
    = \left( r_{\rm out} / c \right)
    \, \left( 1 + ( 2 \slash 3) \, \sin{i} \right)$,
where $i$ is the inclination angle between the accretion disc, with $i=0^{\circ}$ being face on and $i=90^{\circ}$ being edge on. In our further analysis, we fit the SED and lag spectrum by fixing $i=0^{\circ}$ and $45^{\circ}$. Therefore, the expected mean lag at the location of the outer disc with the best-fit $r_{\rm out}$ is $\left<\tau\right>_{\rm rim}\approx 120$ and 180 days, for $i=0^{\circ}$ and $45^{\circ}$, respectively. Inside the steep rim, the flat disc with $T_{\rm eff}\propto r^{-3/4}$ produces a relatively prompt response with a lag increasing as $\tau_{\rm disc}\propto \lambda^{4/3}$. The flat disc response dominates at UV and blue optical wavelengths, and the rim response becomes increasingly prominent at redder optical wavelengths, as shown in Fig.\,\ref{fig:steep}. The model lag spectrum with $\epslp=1$ provides a good fit to the observed continuum lag. However, the small near-infrared excess around $\sim 1.5~\rm \mu m$ is caused by the black body emission at the rim where $T_{\rm eff} \sim 5000~\rm K$. 

The Appendix Fig.\,\ref{fig:bowl_cov} shows the corner plot of the accretion disc and lamp-post parameters for the MCMC iterations. The marginalized posterior distribution of the power disc shape parameter, $\log k\equiv \log ({\rm d} \ln H \slash {\rm d} \ln R)$, flattens as $\log k \gtrsim 2$ because the `rim' becomes so steep that a further increase of the power law slope has little effect on the SED and continuum lag. The MCMC fit yields an inner radius $r_{\rm in}\sim r_\text{g}$, close to the radius of the ISCO for a maximum-spinning black hole. As shown on the lower right panel of Fig.\,\ref{fig:steep}, this behavior is due to the discrepancy between the observed \textit{Swift} band SED and the model, where the model falls below the observed extinction-corrected SED, requiring a high temperature, $T\ga10^5$\,K, on the inner part of the disc. However, our model neglects relativity effects near the black hole, and assumes that the SED is dominated by the blackbody emission from the disc. In the inner region of the disc, the contribution due to non-thermal emission becomes more significant, and it becomes necessary to consider the `warm Comptonising' component in addition to the blackbody emission when fitting the UV part of the SED \citep{Kubota2018}. While these approximations render the inner-disc parameters and the UV end of the SED less reliable, our focus is on understanding how the shape of the blackbody reprocessing surface in the outer disc and rim can affect both the observed time delay and SED.
In summary, with the caveats noted above, this `flat disc + steep rim' model 
provides a useful parameterisation that
achieves a plausible fit simultaneously to both the lags and the SED. 

\subsection{Dust in the accretion disc and the connection to the BLR} \label{frado}

There is great uncertainty about how far accretion discs extend outwards in AGN. In particular, the question as to whether accretion discs extend out to cool enough regions for dust to form and survive is unresolved \citep{Landt2023}. This has important implications for the standard picture of AGN structure and the formation of the BLR, the origins of which are currently unknown. One formation mechanism introduced by \citet{Czerny2011} describes the BLR as originating from dusty outflows from the accretion disc in a mechanism referred to as FRADO \citep[`Failed Radiatively Accelerated Dust-driven Outflow;][]{Czerny_2015}. Under this model, the inner edge of the BLR occurs near an effective temperature of $T_{\rm eff}\sim1000$~K, below which dust can survive in the disc atmosphere. 
This agrees with observations of 35 AGN (taken from the \citet{Bentz2009} sample) by \citet{Czerny2011}, which found the BLR of all sources resided at locations corresponding to an accretion disc temperature of approximately $1000$~K. 

\begin{figure}
    \centering 
    \includegraphics[width=\columnwidth]{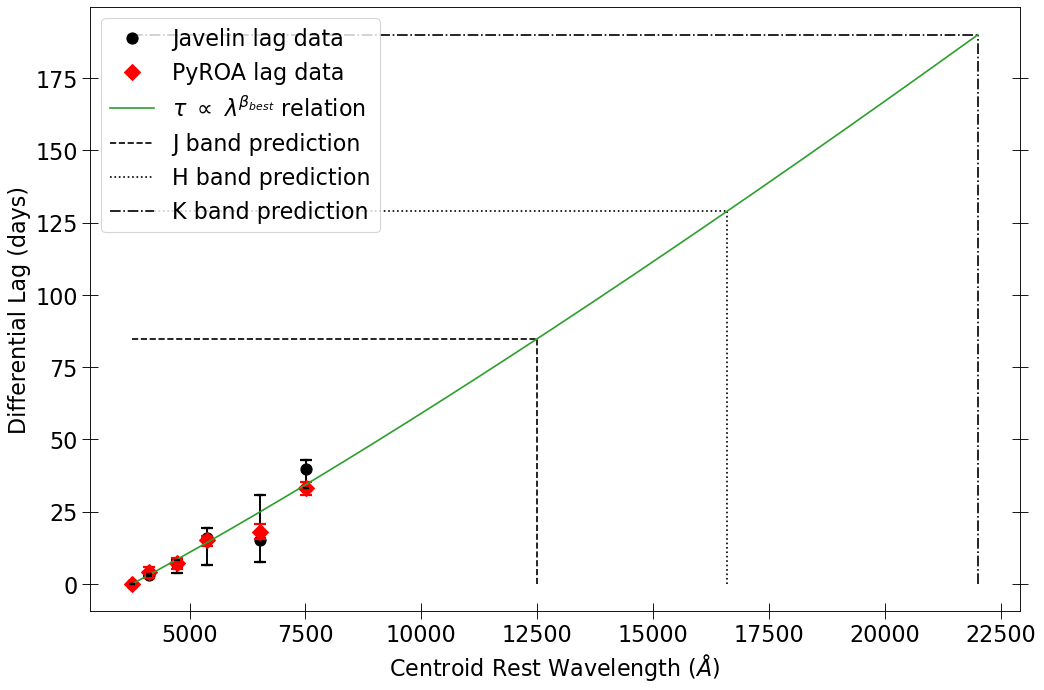}
    \caption{Extrapolation of the $\tau\propto\lambda^{\beta}$ power law fit to the combined \texttt{PyROA} and \texttt{Javelin} RM results into the near-infrared $JHK$ regime. This allowed the lag predictions in Table~\ref{table:JHK} to be made.}
    \label{fig:JHK}
\end{figure}

To test whether this relationship holds true for 3C\,273, we extrapolate our optical lag measurements into the near-infrared, thus estimating the lag for hypothetical dust-forming regions of the accretion disc, and compare the result with measurements for the BLR. We do this by extrapolating the fitted power laws obtained from a combined fit to the lags of both the \texttt{Javelin} and \texttt{PyROA} results shown in Fig.\,\ref{fig:tau-lam} to the $\sim 1000$~K region of the accretion disc. 
We estimate that this dusty region emits most strongly at the observed wavelengths of the $J$ (1.235\,$\mu$m), $H$ (1.662\,$\mu$m) and $K$~bands (2.159\,$\mu$m) and have then extrapolated the fit to these wavelengths (Fig.~\ref{fig:JHK}). 

The lag estimates derived from both the theoretically motivated power law ($\beta = 4/3$) and from the best-fit power law ($\beta = 1.17 \pm 0.25$) are given in Table~\ref{table:JHK}. We have converted to rest-frame absolute values using 
the average of $\tau_0 = 19.7$~days from the estimated $B$-band rest-frame reference lags, excluding the $u$ band, as listed in the last column of Table~\ref{table:lags}, which are required to bring the observations in agreement with theoretical predictions. 
The resulting near-IR lag estimates are in the range of \mbox{$\sim 100 - 200$~days} and consistent with the location of the BLR of $145\pm35$~days as measured from near-IR interferometric observations by \citet{Gravity2018}. This suggests that the BLR in 3C\,273 may form from dust-driven mechanisms in agreement with the FRADO model. Our results also indicate that the accretion disc in 3C\,273 may extend outwards to dusty regions thus challenging the standard AGN paradigm by suggesting there may be a continuous transition from the disc to the dusty torus. Our extrapolated near-IR lags are also consistent with the location of the outer edge of the disc at $r_{\rm out}\sim 150$~light-days, which we obtained for the disc model with a steep rim (Section\,\ref{rim}). A dedicated near-IR RM campaign on 3C\,273 will be able to probe these putative dusty accretion disc regions. Such a campaign has been recently completed by some of us and the results from it are forthcoming.  


\begin{table}
\begin{center}
\caption{Extrapolated lag predictions for the $J$, $H$ and $K$ band emitting regions of the accretion disc in 3C\,273 from the combined \texttt{PyROA} and \texttt{Javelin} RM results. 
The differential lag predictions are in observed frame and were converted to rest-frame absolute values using the average $B$ band reference lag of $\tau_0 = 19.7$~days, which is required to bring observations in agreement with theoretical predictions.}
\label{table:JHK}
\begin{tabular}{||c | c | c | c | c | c ||} 
 \hline
 Filter & $\lambda_{\rm eff}$ & \multicolumn{2}{c|}{Differential Lag Predictions} & \multicolumn{2}{c|}{Absolute Lag Predictions}\\
 & (\AA) & \multicolumn{2}{c|}{(days)} & \multicolumn{2}{c|}{(days)}\\
 \cline{3-6}
 & & \emph{$\beta=\frac{4}{3}$} & \emph{$\beta \sim 1.17$} & \emph{$\beta=\frac{4}{3}$} & \emph{$\beta \sim 1.17$} \\
 \hline
 J & 12350 & 69 & 54 & 79 & 67 \\
 \hline
 H & 16620 & 113 & 86 & 117 & 94 \\
 \hline
 K & 21590 & 170 & 125 & 166 & 128 \\
 \hline
\end{tabular}
\end{center}
\end{table}

\subsection{On the optical contamination from the radio-loud jet} \label{jet}

The synchrotron emission from loud radio jets in AGN extends over almost the entire electromagnetic spectrum and could be a potential source of contamination in RM campaigns. This is especially important for blazars like 3C\,273 which are thought to have relativistically beamed jets lying on or close to the our line of sight, giving them much greater apparent luminosities \citep{1997iagn.book.....P}. Our FVG analysis (Section~\ref{FVG and jet}) found the spectrum of the variable component of our data to be consistent with an accretion disc $F_\nu\propto\nu^{1/3}$ power-law, which suggests that variability in our light-curves is dominated by the accretion disc. This makes it unlikely that the jet or any other source of contamination (e.g. diffuse continua from the BLR) would significantly affect the shape of our light-curves to bias the RM results. There is a hint of the rms SED in Fig~\ref{fig:SED} increasing slightly towards the near-IR which could weakly suggest jet activity. However, this is statistically insignificant relative to the errors. This agrees with the findings of \citet{Yuan_2022} who, by fitting blackbody and synchrotron emission models to the SED of 3C\,273, found the optical continuum emission (at the rest wavelengths they observed  $\leq 5500$\AA) to be dominated by the accretion disc with an insignificant jet contribution. \citet{Yuan_2022} detected quasi-periodic variability for this accretion disc emission on a timescale of $\sim 3.39 \pm 1.13$ years. This is in broad agreement with the $\sim 3$ year quasi-periodic variability evident in our light-curves in Fig\,\ref{fig:lc} which we fit with sine curves in Section~\ref{fast_slow}. This further indicates that the only significant variability present in our light-curves originates from the accretion disc. 

\citet{1998A&A...340...47P} decomposed optical and UV light-curve data from 3C\,273 into short-timescale relatively blue 
and long-timescale relatively red 
components, which they attribute to the accretion disc and jet, respectively. A similar attempt in the optical was done by \citet{Li_2020} through modelling the disc and jet variability as DRW’s. Both studies estimated the timescale of their modelled jet variability to be upwards of $\sim10$ years which means we can expect to detect no significant jet variability over our $\sim 3$-year long data-set. Whilst \citet{Li_2020} estimate the jet to contribute $\sim 10-40\%$ of flux in the optical, as it transitions from its quiet to loud states, this should be present as a constant background source over the time frame of our observations and so is unlikely to impact our RM results.

\subsection{Fast and slow variations} \label{fast_slow}

Several studies \citep[e.g.][]{Uttley_2003,Breedt_2009} have observed optical continuum variability in AGN with both short and longer timescale components. The faster variability can usually be attributed to X-ray reprocessing but the existence of a slower component may suggest an additional source of variability. These two variabilities in AGN accretion discs were investigated in detail by \citet{JVHS2020} for the AGN Fairall\,9 using high-quality optical light-curves spanning a very large range in wavelength (from $\sim 2000 - 9000$~\AA). Their continuum light-curves displayed rapid variability that was fit in a RM procedure. The lags associated with this faster variability increased with wavelength as expected for X-ray reprocessing in the disc. However, the light-curves also had a slow variation (>100 days) which they fit with quadratic functions of time to estimate the lags between them. The longer-timescale lags broadly decreased with wavelength, a behaviour consistent with temperature fluctuations moving inwards through the disc, although the measured time lags were too short for such fluctuations to be the fully accepted interpretation for them. 

\begin{figure}
    \centering 
    \includegraphics[width=\columnwidth]{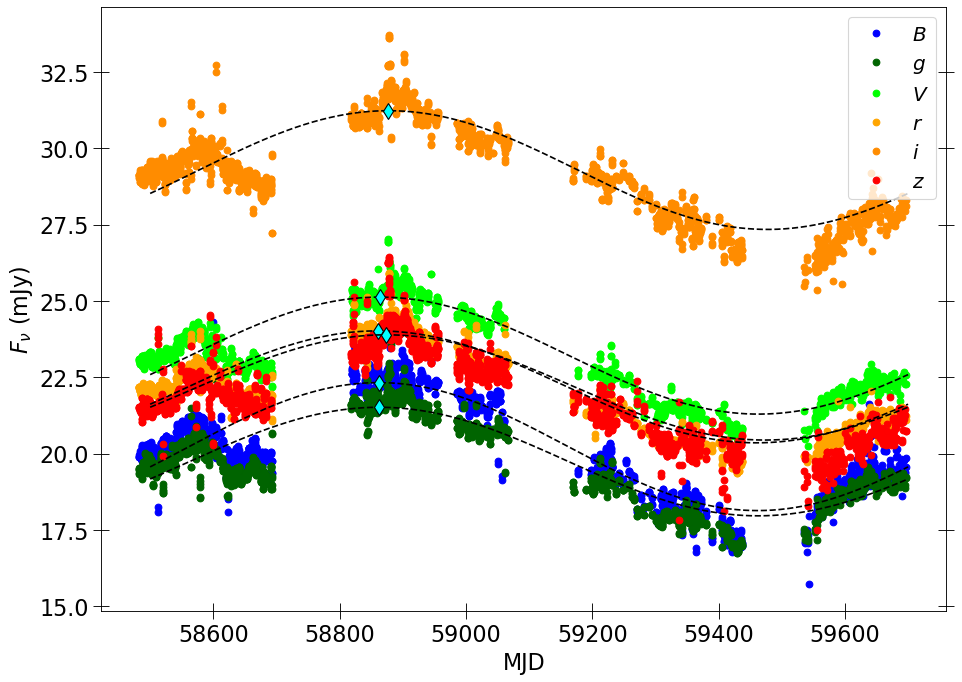}
    \caption{LCO light-curves each fitted with a sinusoidal function. The light-curves have been colour-coded according to their photometric band. For reference, the maximum of each sine wave is also plotted (diamonds).}
    \label{fig:trend1}
\end{figure}

The light-curves obtained in our observing campaign of 3C\,273 also exhibit a quasi-periodic trend, indicative of a longer timescale (of $\sim 3$ years) variability in addition to the variability fit in our RM analysis (see Fig.\ref{fig:trend1}). But we caution that, since we observe less than one period, we cannot exclude that the apparent periodicity is introduced by red noise \citep{Vaughan2016}. In order to establish true periodicity in variations would require a monitoring campaign long enough to cover multiple variability cycles. 

\begin{figure}
    \centering 
    \includegraphics[width=\columnwidth]{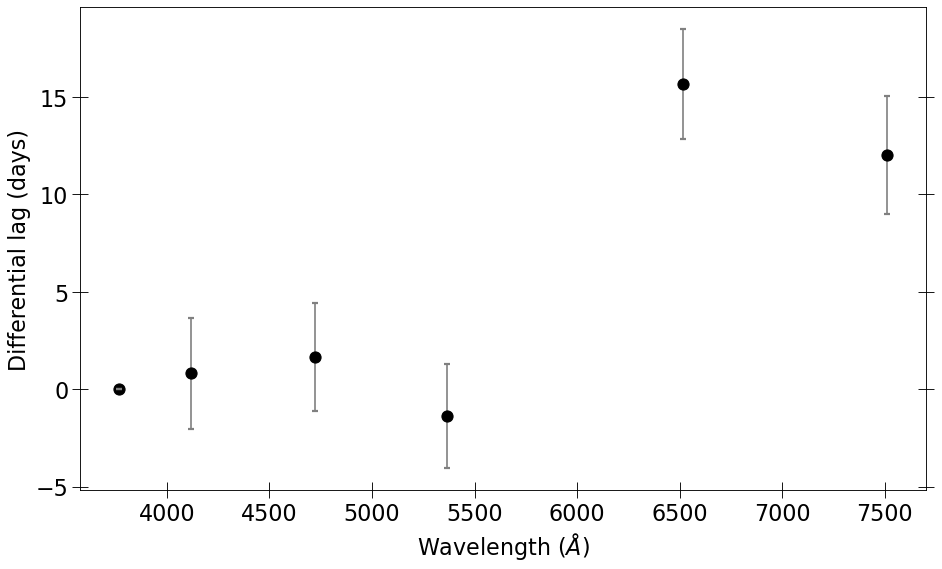}
    \caption{The lag spectrum for the fitted trends in Fig~\ref{fig:trend1}. Lags are relative to the $B$ band.}
    \label{fig:trend2}
\end{figure}

We have estimated the lags associated with the slower apparent variability following the approach of \citet{JVHS2020}. But, owing to the more complex shape of the trend, we cannot use the quadratic expression used by them and instead fit the trend with a sinusoidal function of the form:

\begin{eqnarray}
\label{fast-slow}
F_{\nu}(\lambda,t) = \bar{F}_{\nu}(\lambda) + A(\lambda)\,\sin{\left(\frac{2\,\pi}{P}(t-t_0(\lambda))\right)},
\end{eqnarray}

\noindent
where $\bar{F}_{\nu}(\lambda)$ is the mean flux of the light-curve at wavelength $\lambda$, $P\approx1000$\,d is the period (set to the length of the observing period) and $A(\lambda)$ and $t_0(\lambda)$ are the amplitude and shift of the sine waves, both fit as free parameters. 
The best-fit sine curves shown in Fig.\,\ref{fig:trend1} have roughly the same amplitude in mJy in all bands. The fitted time shift $t_{0}(\lambda)$, relative to that for the $B$ band, is shown in Fig~\ref{fig:trend2}. For the limited wavelengths in common with the data of \citet{JVHS2020}, we find a very similar trend (see top panel of their Fig.~7); a broadly decreasing lag with wavelength for wavelengths $\lambda \la 6000$~\AA, followed by a sudden jump to a higher lag value and then again a continued decrease of the lags with wavelength. But, in contrast to the findings of \citet{JVHS2020} for Fairall~9, the jump in 3C~273 is to much higher lag values. 


\section{Summary and Conclusions} \label{Conclusion}

This investigation reports results from the first accretion disc reverberation mapping campaign on the quasar 3C\,273. This is based on high-cadence light-curve data spanning the near-UV to the near-IR/optical collected by the LCO over a $\sim3.3$-year long monitoring campaign, together with contemporaneous UV fluxes from $Swift$ of a coarser cadence. Our main results can be summarised as follows:

\vspace*{0.2cm}


(i) Fits to the light-curves by the reverberation mapping algorithms \texttt{Javelin} and \texttt{PyROA} are used to obtain two sets of differential lag estimates. Both sets of results agree with each other within their uncertainties and give lags 
a factor of $\sim 2-3$ and a factor of $\sim 4-7$ larger than predicted from the thin disc model, assuming a dimensionless Wien scaling factor of $X=4.96$ and $X=2.49$, respectively. Therefore, 3C\,273 displays the `accretion disc size problem' widely observed for low-luminosity AGN.

(ii) Physically motivated power-laws of the form $\tau\propto\lambda^{\beta}$ were fit to the \texttt{Javelin} and \texttt{PyROA} lag estimates for the Year~3, which showed the largest variability. Allowing $\beta$ to vary generated fits in all cases which reproduced the thin disc model expectation \mbox{$\beta = 4/3$} within uncertainty limits. Power-law fits to a contemporaneous optical spectrum of 3C\,273 produced a $\beta$ estimate also consistent with the thin disc model and previous observations. Therefore, our findings indicate that the accretion disc in 3C\,273 conforms to the thin disc model. 

(iii) We test the `flat disc with a steep rim' model of \citet{Starkey2023} and find that it can fit simultaneously the inter-band lags and the SED of the spectral variations in 3C\,273. The best-fit model places the outer edge of the disc at $R_\text{out}\sim 150$~light-days, with a relatively steep outer rim with a height of $\sim 0.6$~light-days. The disc temperature profile scales as $T\propto r^{-3/4}$, falling to  $1000$\,K  and then rising to $5000$\,K at the irradiated crest of the rim.

(iv) Extrapolating  the $\tau(\lambda)\propto \lambda^{4/3}$ power-law fit to our optical lags into the near-IR regime, we predict the lags for the putative dusty outer regions of the disc of $\sim 100 - 200$~days. These lags are in good agreement with near-IR interferometric measurements of the BLR radius as well as the best-fit outer edge for the `flat disc with a steep rim' model. Therefore, dust-driven outflow models for the formation of the BLR, which suggest that the inner edge of the BLR should occur where the disc atmosphere temperature crosses the $\sim 1000$~K dust sublimation temperature, might be a viable proposition. A more rigorous test of these models would require a dedicated near-IR RM campaign on 3C\,273.     

(v) A flux variation gradient analysis and the structure function both show that the variability in our light-curves is dominated by the accretion disc. This validates the RM lag estimates we obtain and implies insignificant contribution of the loud radio jet to the variability over our observing period. This is in broad agreement with the findings of \citet{Yuan_2022}, who found an insignificant jet contribution to the optical fluxes in 3C\,273 and estimated an accretion disc variability timescale broadly consistent with the long timescale trend evident in our light-curves.


\section*{Acknowledgements}

We thank Jonathan Gelbord for making the processed archival {\it Swift} data available. JPT acknowledges the support of Science and Technology Facilities Council (STFC) studentship ST/X508433/1. HL acknowledges a Daphne Jackson Fellowship sponsored by the STFC and support from STFC grants ST/P000541/1, ST/T000244/1 and ST/X001075/1. JVHS and KH acknowledge support from STFC consolidated grant to St. Andrews (ST/M001296/1).

\section*{Data Availability}

The raw data underlying this work can be downloaded from the LCO archive at \mbox{http://archive.lco.global} and the {\it Swift} archive at \mbox{https://www.swift.ac.uk/}. The processed data underlying this work are available on reasonable request from the authors.



\bibliographystyle{mnras}
\bibliography{3c273reverb}




\appendix

\section{Outlier Rejection Using Structure Function (SF)}
We note that the flux error assigned to outliers in the light curves can sometimes be small after the \texttt{PyROA} inter-calibration procedure (See Fig.\,\ref{fig:lc}). These outliers cause overfitting of the light curves when the ROA parameter $\Delta$ becomes small, so $X(t)$ follows rapid variations in the data and picks out the outliers as part of the variation signal. To minimize the effect of thees outliers on the fitted light curves, we can either enlarge the error bars or remove these outliers. An efficient way of outlier detection is to use structure function (SF) analysis.

With the fitting procedure described in $\S$\ref{structure}, we first fit the variance of the SF for each light curve with the DRW model given by Eq.\,\ref{eq:drw}. For each light-curve datum $D_i$, we compute $\chi_i^2$ relative to the fitted SF model, where $\chi_i^2$ is evaluated by summing over the individual fit residuals for each of the $i, j$ pairs within the proximity ($\Delta t_{ij}\leq 10$\,days) of point $i$:
\begin{equation}
    \chi_i^2 = \sum^{N-1}_{\substack{j \\ (\Delta t_{ij}\leq 10 \,{\rm days})}} \frac{(D_i-D_j)^2}{\sigma^2_i + \sigma^2_j + SF(t_i - t_j)},
\label{eq:sfchi2}
\end{equation}
where $D_i$ and $D_j$ correspond to the $i$- and $j$-th value in the structure function computed with Eq.\,\ref{eqn:SF}. We reject data with $\chi_i^2$ greater than a threshold $(\chi_i^2)_{\rm threshold}$. An example of using this algorithm is shown in Fig.\,\ref{fig:sf_rej}. As a demonstration, only two iterations is applied for the outlier rejection. The upper panel shows the $u$-band light curve after inter-band calibration. The gray error bars are the rejected outliers from the first iteration, and the red error bars correspond to the rejected points with $\chi_i^2 > (\chi_i^2)_{\rm threshold} = 70$. The final outlier-rejected light curve is shown in blue. The reason for computing $\chi_i^2$ with a proximity of 10 days around the $i$-th point is because of the large fluctuation in the light curves across epochs. This means that near the maxima or minima of the light curves, the sum of $(D_i-D_j)^2$ becomes larger than that when computed near the mean of the light curve data. If we sum the individual fit residual across the full light curve, the points near the light curve extrema will have a high value of $\chi_i^2$ and will be incorrectly removed as an outlier. 

The above step is repeated by reevaluating the $\chi_i^2$ values after rejecting each outlier by assigning the rejected data an infinitely large error bar so they do not contribute to $\chi_i^2$.

\begin{figure}
    \centering 
    \includegraphics[width=\columnwidth]{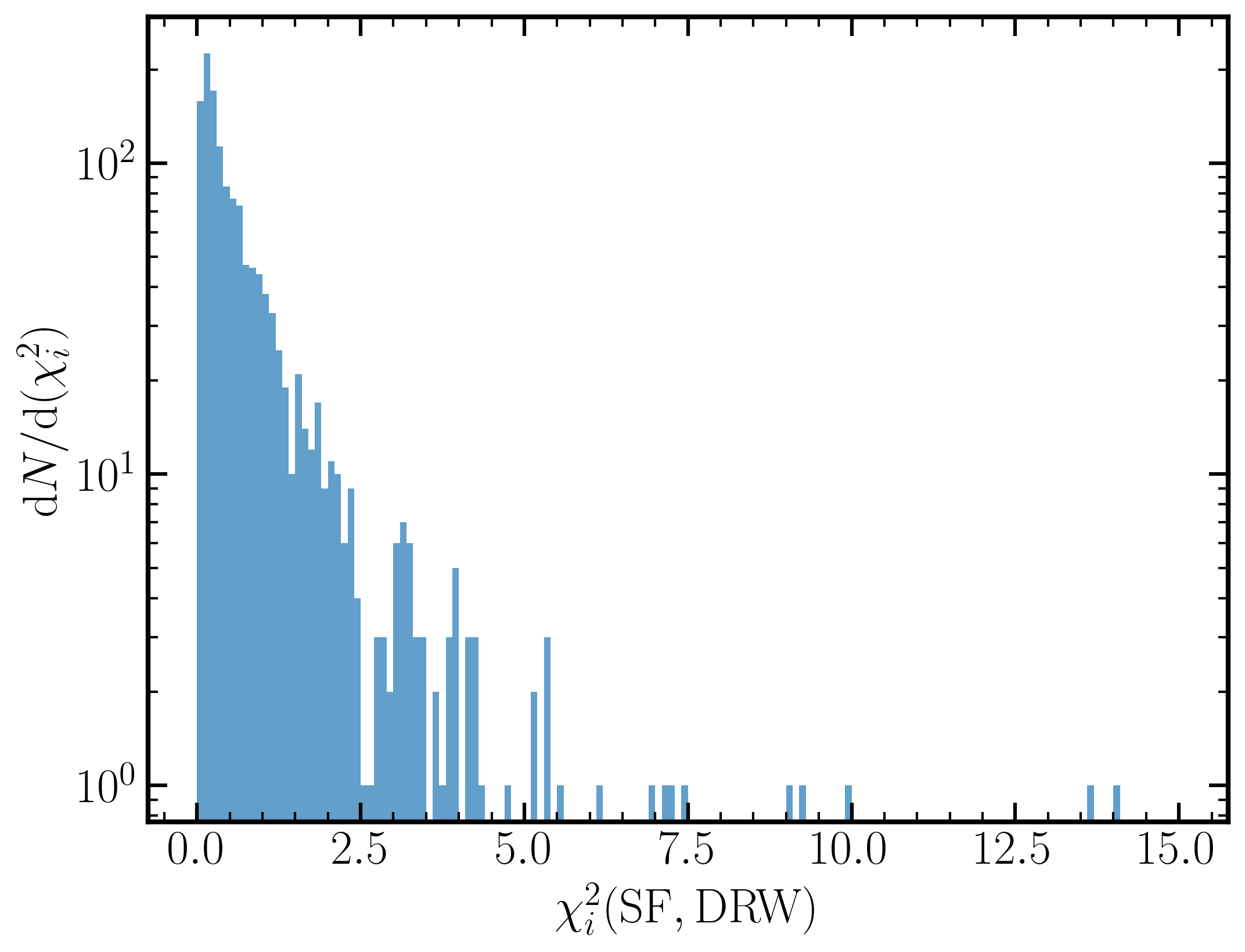}
    \caption{Distribution of $\chi_i^2$ from the SF analysis for the $u$-band light curve. Here $\chi_i^2$ is computed for each of the data on the light curve with Eq.\,\ref{eq:sfchi2}. All points with $\chi_i^2$ larger than a certain threshold are regarded as outliers.}
    \label{fig:sfchi2}
\end{figure}

\begin{figure}
    \centering 
    \includegraphics[width=\columnwidth]{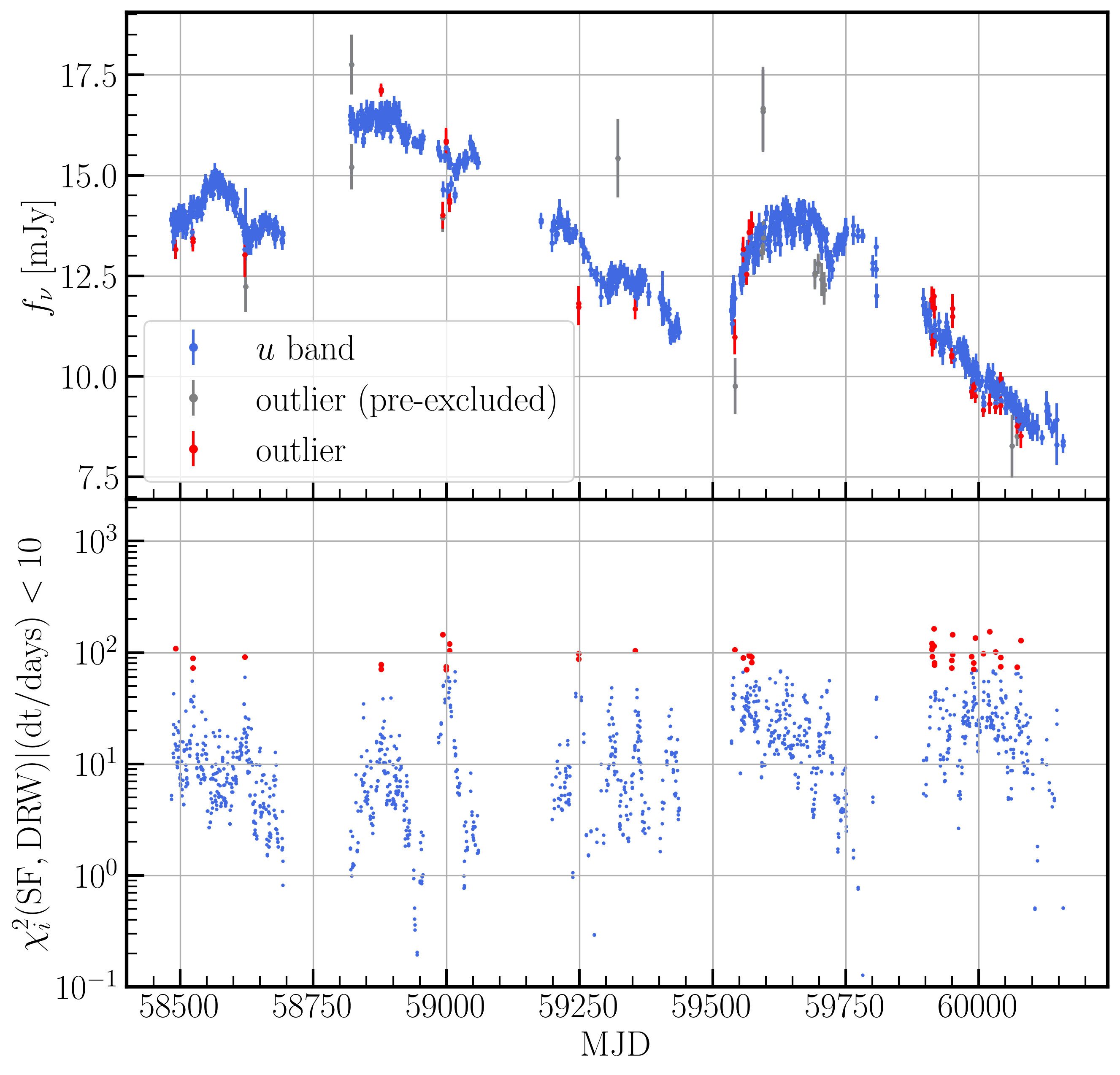}
    \caption{A demonstration of the outlier rejection algorithm with two iterations. \textbf{\textit{Top:}} The $u$-band light curve after inter-band calibration. The grey data points are the rejected outliers from the first iteration. In the second iteration, the grey error bars are assigned with infinitely large error bars, so they do not contribute to the $\chi_i^2$ calculation. The red error bars correspond to the rejected points in the second iteration. The final outlier-rejected light curve is shown in blue. \textbf{\textit{Bottom:}} Computed $\chi_i^2$ values (with $\Delta t_{ij}\leq 10$ days) in the second iteration for each data point on the $u$-band light curve, plotted as a function of time. The red scatter corresponds to the rejected points with $\chi_i^2 > (\chi_i^2)_{\rm threshold} = 70$.
    }
    \label{fig:sf_rej}
\end{figure}

\section{Additional figures}


\begin{figure*}
    \centering 
    \includegraphics[width=1\linewidth]{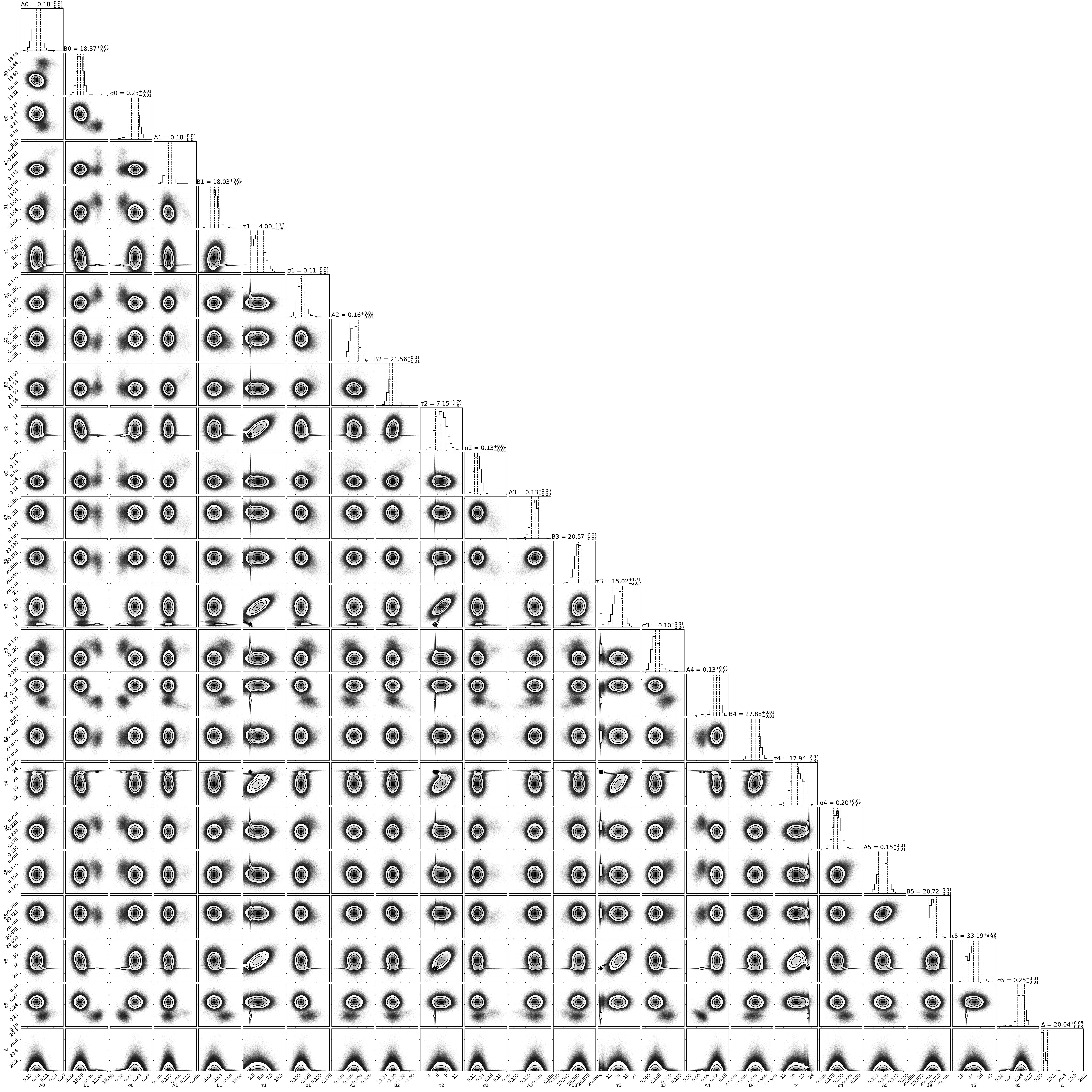}
    \caption{\texttt{PyROA} corner plots to the 3C\,273 fits in Fig.~\ref{fig:pyroa_fit} and lag results in Table~\ref{table:lags}.}
    \label{fig:pyroa_cov}
\end{figure*}

\begin{figure*}
    \centering 
    \includegraphics[width=1\linewidth]{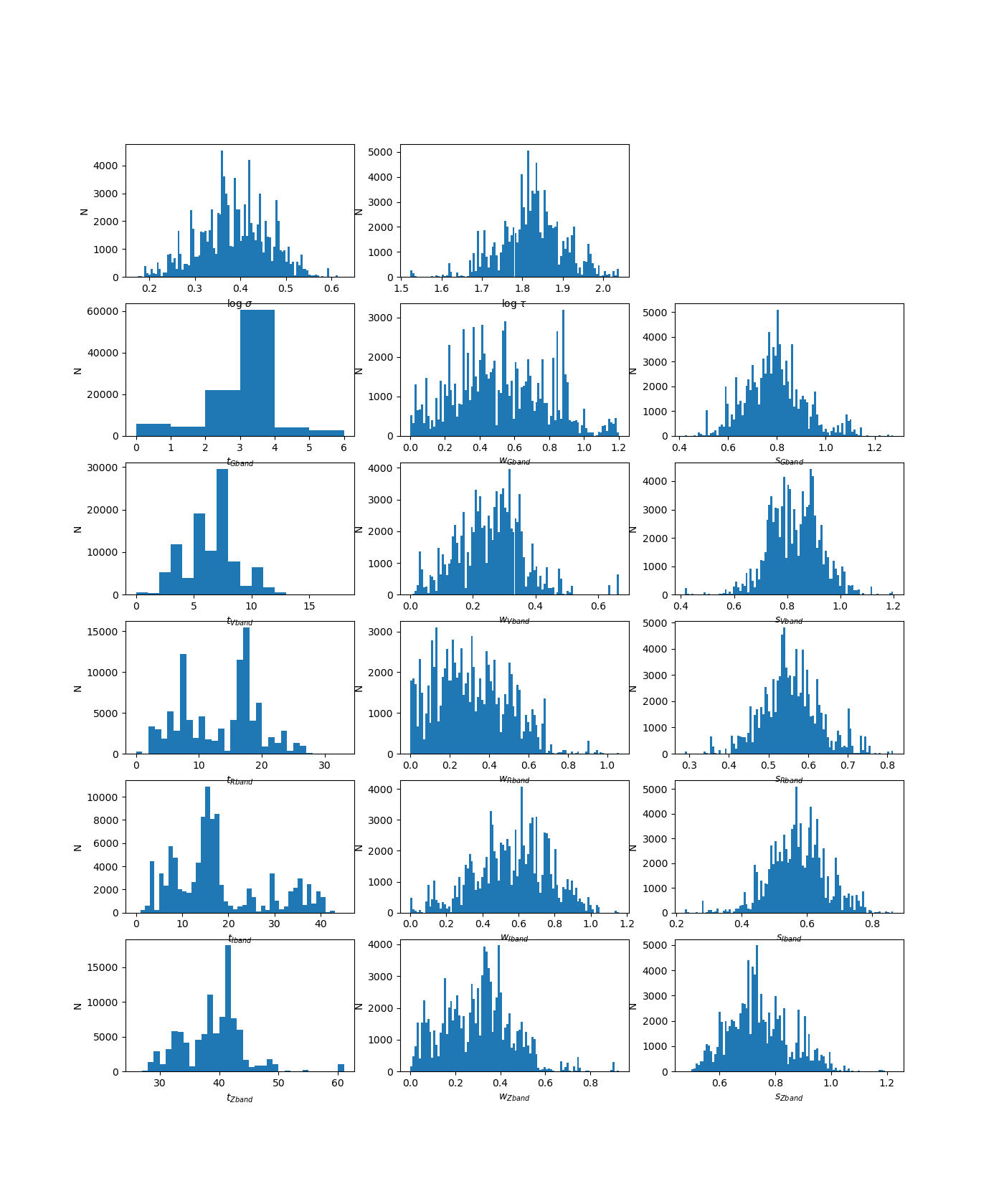}
    \caption{Posterior distributions obtained by \texttt{Javelin} for the fit in Fig.\,\ref{fig:javelin_fit} using a total of $10^5$ MCMC samples. 
    The distributions are for the decorrelation timescale $\tau_{d}$ and short timescale variability amplitude $\hat{\sigma}$ and then the lag $\tau$, tophat width $w$ and scale $s$ for each light-curve.
    \label{fig:jav_par}}
\end{figure*}

\begin{figure}
    \centering 
    \includegraphics[width=\columnwidth]{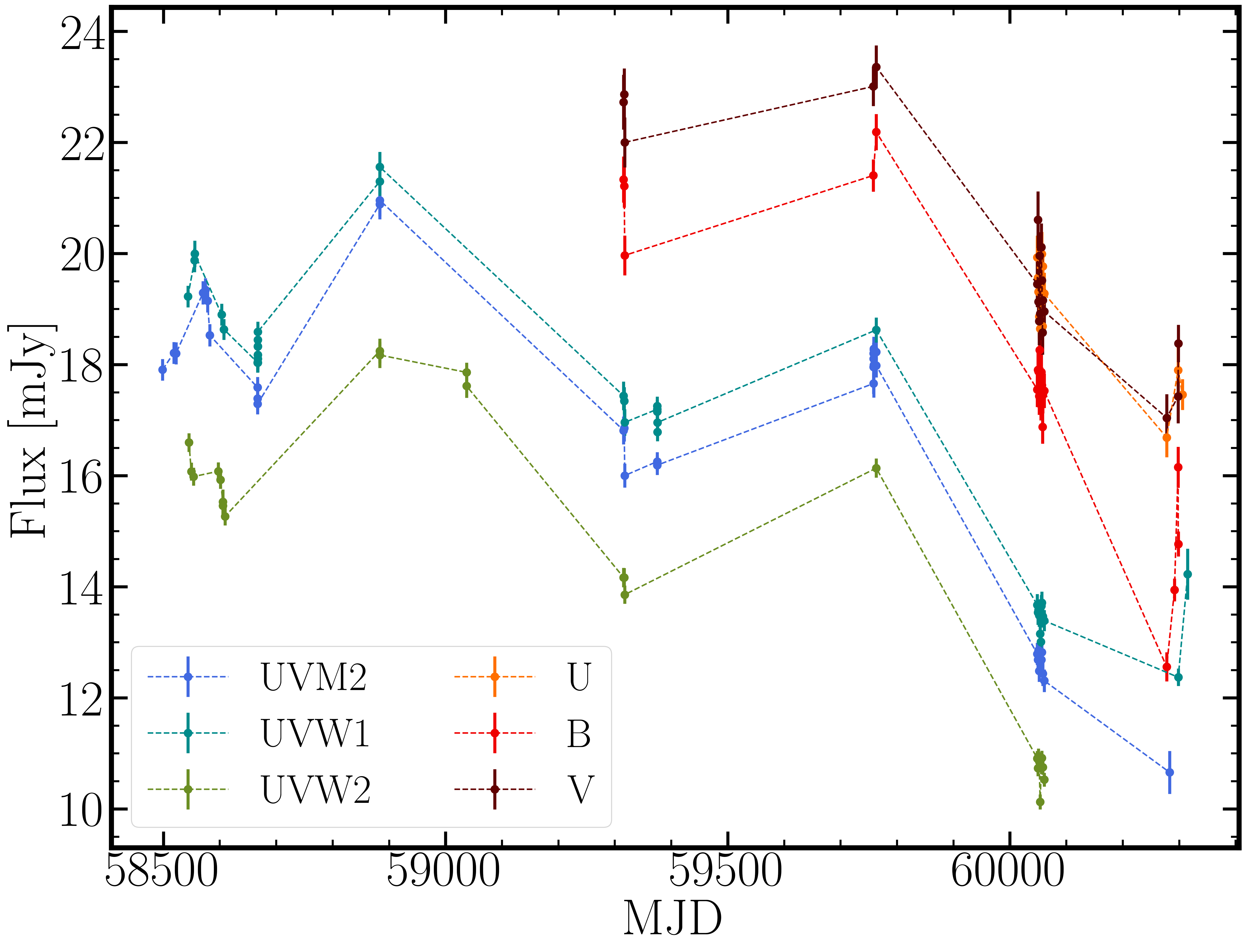}
    \caption{\textit{Swift} UV/optical light curve for 3C\,273.}
    \label{fig:uv_lc}
\end{figure}

\begin{figure*}
    \centering 
    \includegraphics[width=1\linewidth]{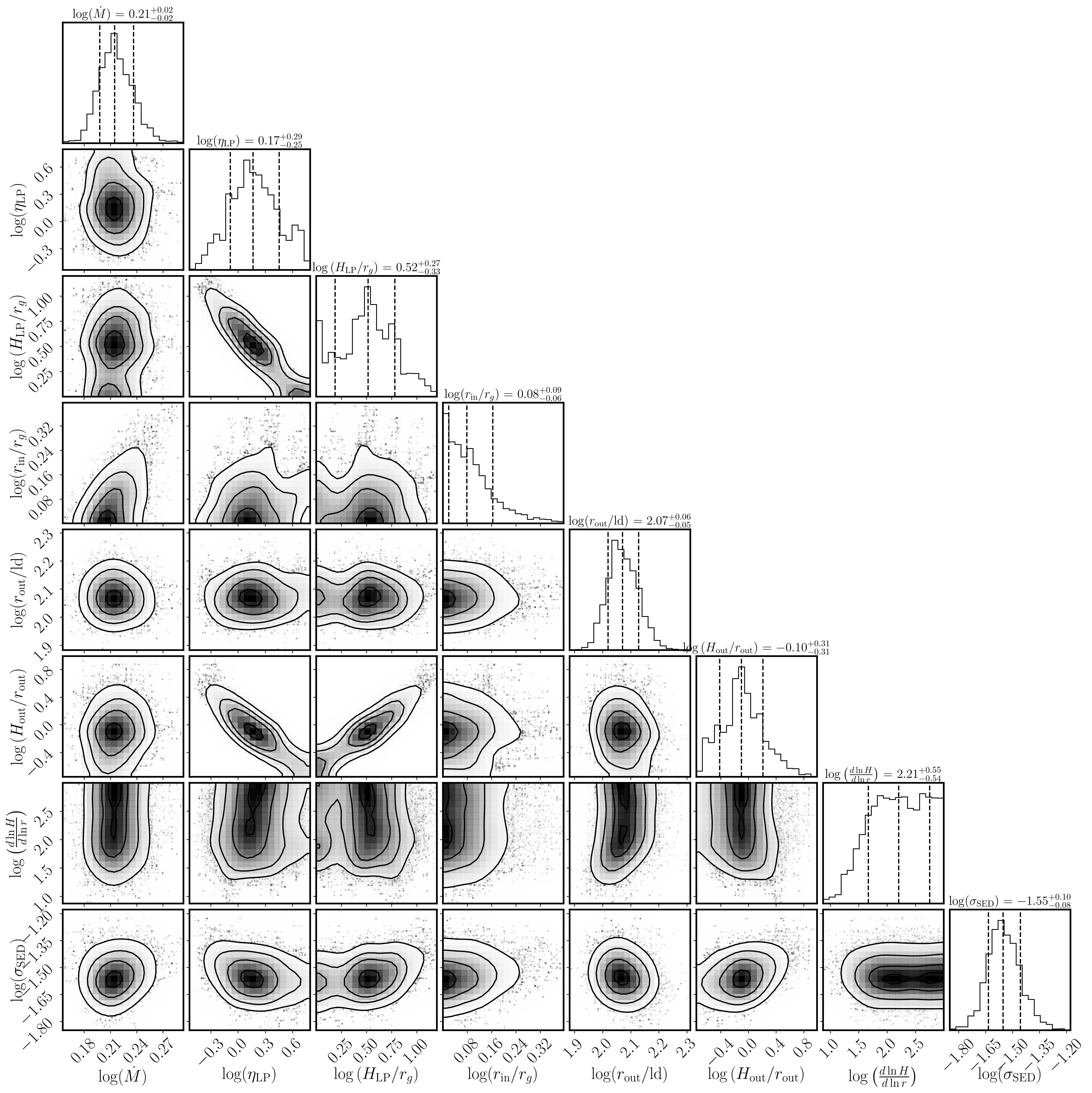}
    \caption{Corner plot of the simultaneously fit (Fig.\,\ref{fig:steep}) to the 3C\,273 SED and time delay spectrum using a power-law shaped accretion disc with fixed inclination $i=0^{\circ}$. The plot shows the distributions of 8 fitting parameters: the accretion rate $\log {\dot M} (\rm M_{\odot}\, yr^{-1})$, lamp-post irradiative efficiency $\log \epslp$, lamp-post height in units of gravitational radius $\log H_{\rm LP}$, inner and outer disc radius $\log r_{\rm in} \,\rm (lightdays)$ and $\log r_{\rm out}\,\rm (lightdays)$, scale height at the outer disc radius $\log H_{\rm out}\slash r_{\rm out}$, power law slope of the disc shape that traces the location of the disc surface $\log k\equiv \log ({\rm d} \ln H \slash {\rm d} \ln R)$, and the flux model uncertainty of the SED $\log \sigma_{\rm SED}$.}
    \label{fig:bowl_cov}
\end{figure*}

\begin{figure*}
\centering
	\begin{tabular}{@{}cccccc@{}}
 \multicolumn{1}{l}{}\\
	\includegraphics[width=\columnwidth]{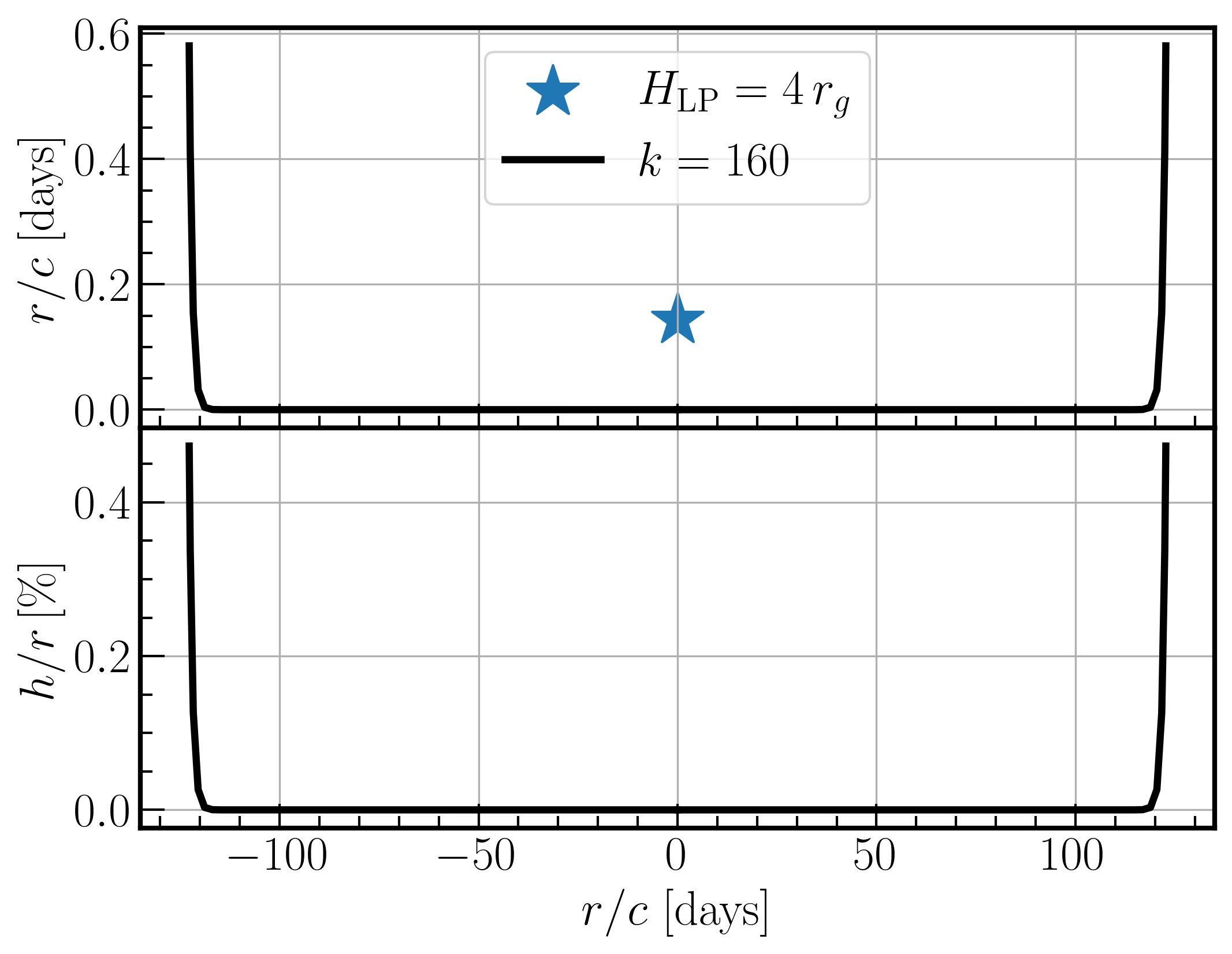}
	\includegraphics[width=\columnwidth]{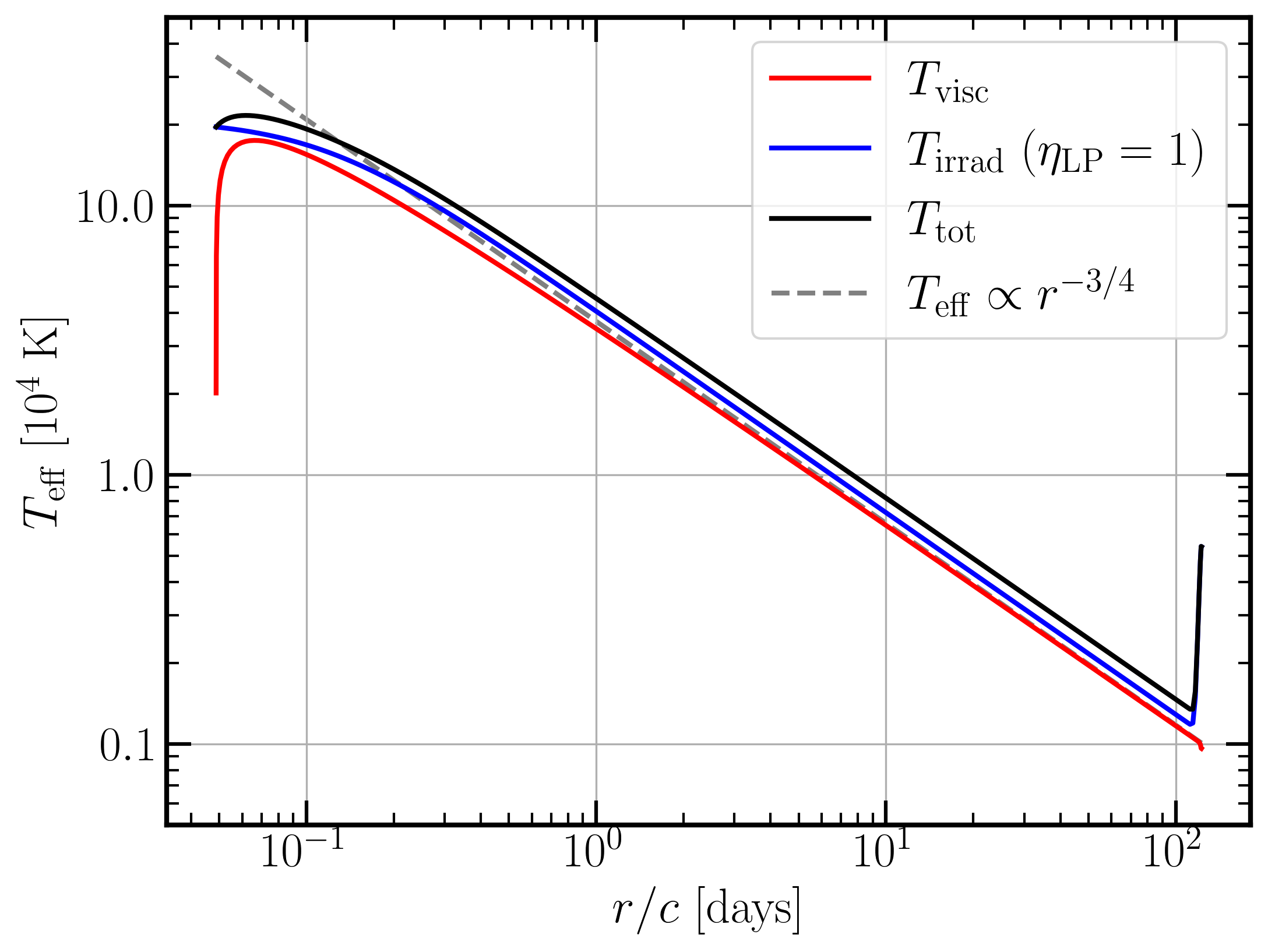}
 \\
  \multicolumn{1}{l}{}\\
	\includegraphics[width=\columnwidth]{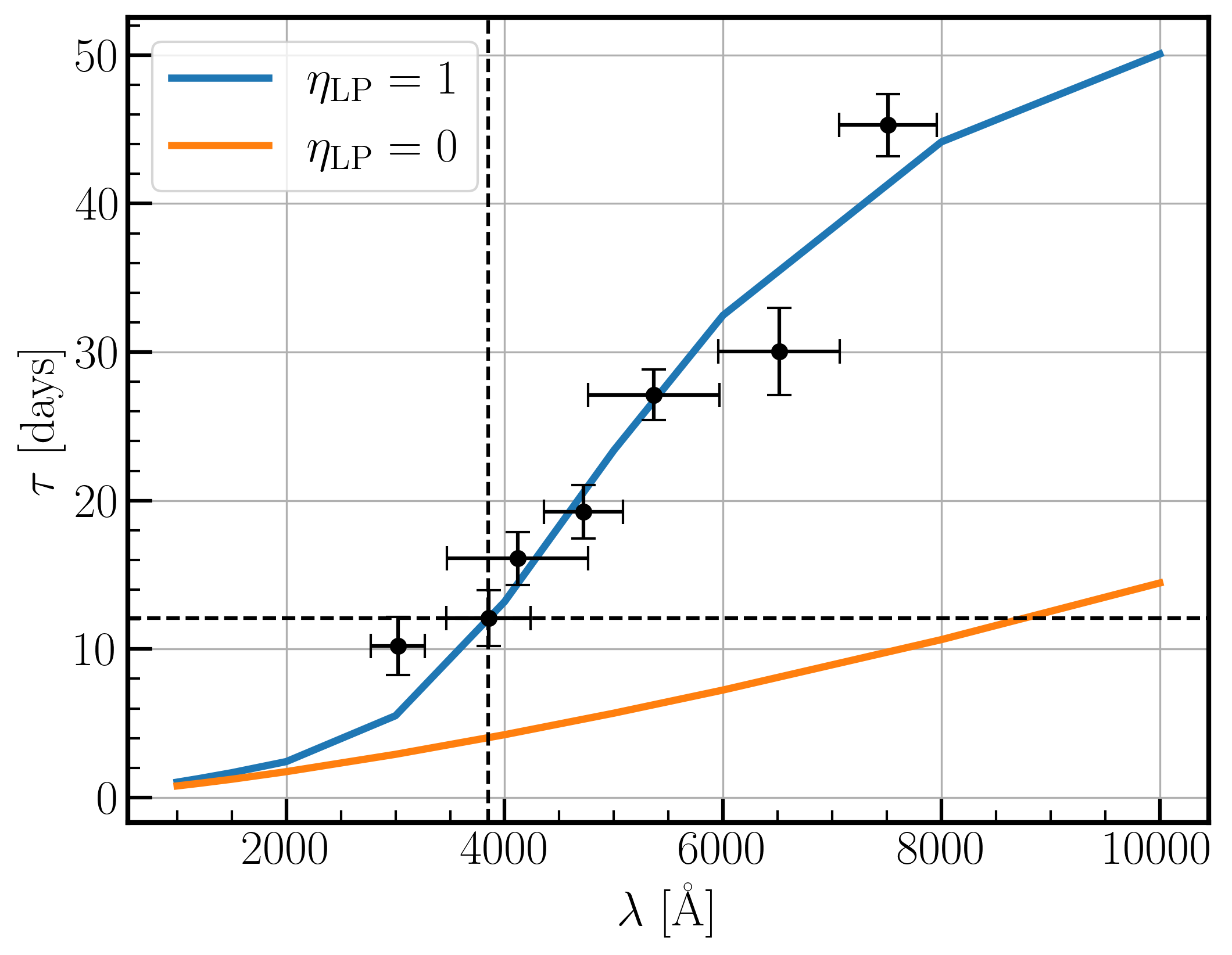}
	\includegraphics[width=\columnwidth]{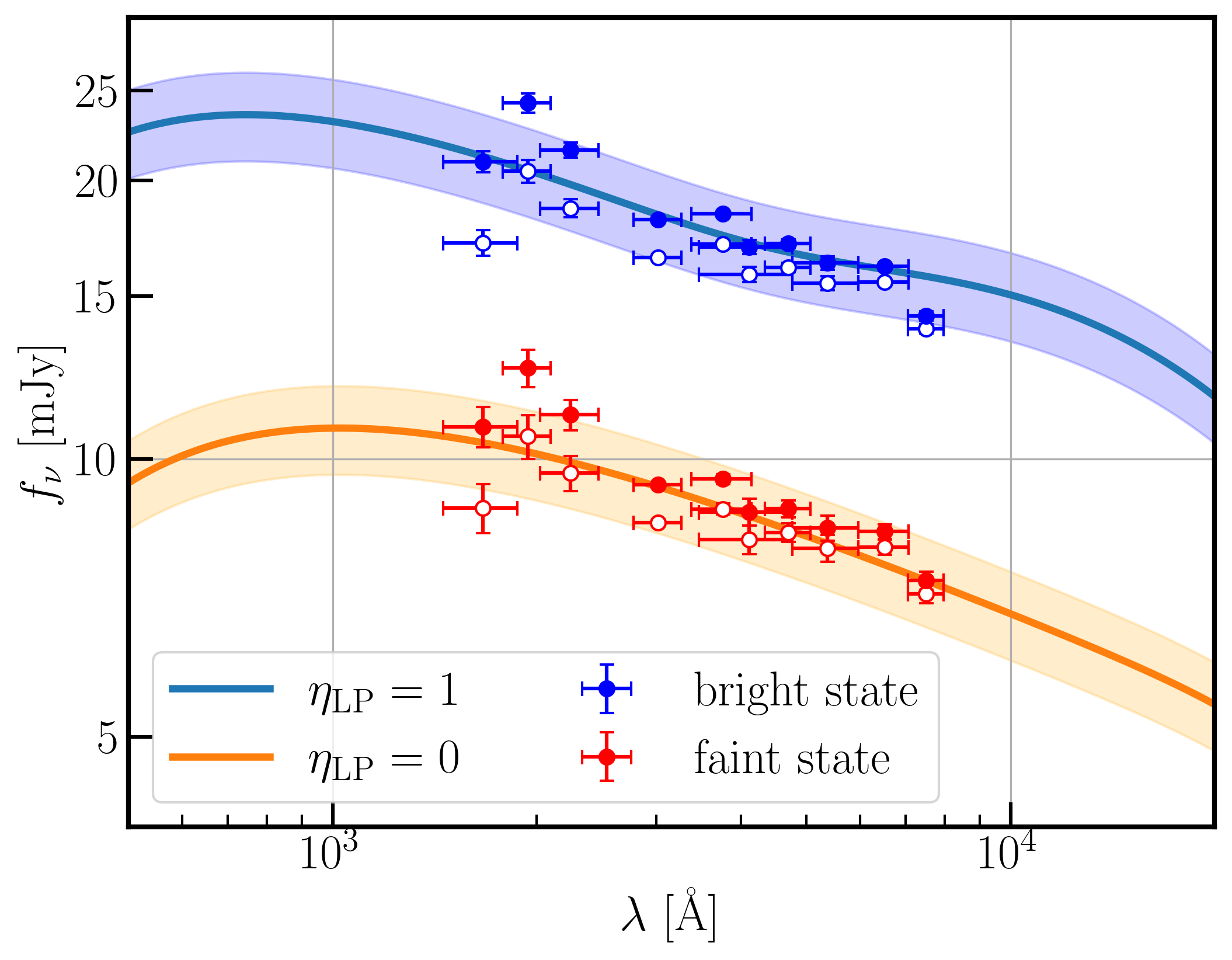}
	\end{tabular}
    \caption{Same as (Fig.\,\ref{fig:steep}), but for a disc inclination of $i=45^{\circ}$. The faint state SED is fitted with $\dot{M} \approx 3.0\, \rm M_{\rm \odot}\, yr^{-1}$.
    }
    \label{fig:steep_45}
\end{figure*}

\bsp	
\label{lastpage}
\end{document}